\documentclass[sigplan,screen,natbib=false]{acmart}

\usepackage[normalem]{ulem}
 \usepackage{algorithmic}
 \usepackage{graphicx}
 \usepackage{textcomp}
 \usepackage{xcolor}
 \usepackage{hhline}
 \usepackage{multirow}
 \usepackage{subcaption}
 \usepackage{url}

\usepackage[symbol]{footmisc}

\AtBeginDocument{%
  }

\copyrightyear{2024}
\acmYear{2024}
\setcopyright{acmlicensed}\acmConference[ASPLOS '24]{28th ACM International Conference on Architectural Support for Programming Languages and Operating Systems, Volume 1}{April 27-May 1, 2024}{La Jolla, CA, USA}
\acmBooktitle{28th ACM International Conference on Architectural Support for Programming Languages and Operating Systems, Volume 1 (ASPLOS '24), April 27-May 1, 2024, La Jolla, CA, USA}
\acmPrice{15.00}
\acmDOI{10.1145/3617232.3624855}
\acmISBN{979-8-4007-0372-0/24/04}

\newcommand{\rev}[1] {\textcolor{black}{#1}}

\begin{document}

\title{GPU-based Private Information Retrieval for \\ On-Device Machine Learning Inference}

%
% The "author" command and its associated commands are used to define
% the authors and their affiliations.
% Of note is the shared affiliation of the first two authors, and the
% "authornote" and "authornotemark" commands
% used to denote shared contribution to the research.

%\numberofauthors{1}

\author{
  % Authors
  \mbox{Maximilian Lam\footnotemark[3]}, 
  \mbox{Jeff Johnson\footnotemark[2]}, 
  \mbox{Wenjie Xiong\footnotemark[4]}, 
  \mbox{Kiwan Maeng\footnotemark[5]}, 
  \mbox{Udit Gupta\footnotemark[7]}, 
  \mbox{Yang Li\footnotemark[2]}, 
  \mbox{Liangzhen Lai\footnotemark[2]}, 
  \mbox{Ilias Leontiadis\footnotemark[2]}, 
  \mbox{Minsoo Rhu\footnotemark[2]}, 
  \mbox{Hsien-Hsin S. Lee\footnotemark[6]}, 
  \mbox{Vijay Janapa Reddi\footnotemark[3]}, 
  \mbox{Gu-Yeon Wei\footnotemark[3]}, 
  \mbox{David Brooks\footnotemark[3]}, 
  \mbox{G. Edward Suh\footnotemark[2]\footnotemark[7]} \\
  \footnotemark[2]Meta AI,
  \footnotemark[3]Harvard University,
  \footnotemark[4]Virginia Tech,
  \footnotemark[5]Penn State,
  \footnotemark[6]Intel,
  \mbox{\footnotemark[7]Cornell University}
}

% By default, the full list of authors will be used in the page
% headers. Often, this list is too long, and will overlap
% other information printed in the page headers. This command allows
% the author to define a more concise list
% of authors' names for this purpose.
\renewcommand{\shortauthors}{Lam et al.}

\begin{abstract}
On-device machine learning (ML) inference can enable the use of private user data on user devices without revealing them to remote servers. However, a pure on-device solution to private ML inference is impractical for many applications that rely on embedding tables that are too large to be stored on-device. 
\rev{In particular, recommendation models typically use multiple embedding tables each on the order of 1-10 GBs of data, making them impractical to store on-device.} 
To overcome this barrier, we propose the use of private information retrieval (PIR) to efficiently and privately retrieve embeddings from servers without sharing any private information. % during on-device ML inference. 
As off-the-shelf PIR algorithms are usually too computationally intensive to directly use for latency-sensitive inference tasks, we 1) propose novel GPU-based acceleration of PIR, and 2) co-design PIR with the downstream ML application to obtain further speedup. Our GPU acceleration strategy improves system throughput by more than $20 \times$ over an optimized CPU PIR implementation, and our PIR-ML co-design provides an over $5 \times$ additional throughput improvement at fixed model quality. Together, for various on-device ML applications such as recommendation and language modeling, our system on a single V100 GPU can serve up to $100,000$ queries per second---a $>100 \times$ throughput improvement over a CPU-based baseline---while maintaining model accuracy.
\end{abstract}





% %% A "teaser" image appears between the author and affiliation
% %% information and the body of the document, and typically spans the


\maketitle
\pagestyle{plain}

%%%%%%%%%%%%%%%

\section{Introduction}
Privacy is an important consideration for real-world machine learning (ML) applications \rev{that use user data}. 
\rev{For privacy-sensitive ML applications, users' demand for stronger privacy protection, as well as  
regulations \cite{gdpr, ccpa} and platform policies \cite{apple_att, google_att}, all increasingly limit the use of private user data. For example, recommendation models, which represent a significant portion of today's ML workloads in practice, inherently rely on individual user data in order to provide personalized recommendations.
Ideally, recommendation systems should provide suggestions to users without revealing private user features even to the service provider.
}
%Recent privacy policies for mobile platforms \cite{apple_att, google_att} limit the type of user data that can be used for server-side computation \cite{apple_att_repercusions}. 

%Violation of data privacy laws at the national level has similarly led to significant repercussions \cite{amazon_data_violation, goog_meta_data_privacy_violation}.

% In context of ML applications like recommendation, increasingly stringent privacy policies pose a significant headwind as it limits the amount of data app makers can access for model inference. Concretely, ML applications like recommendation typically have users' raw data sent to the cloud where it is processed by server-side models \cite{deeprecsys, dlrm}. Increasingly stringent privacy policies limit or altogether eliminate the ability of centralized servers to aggregate user data for ML inference. To overcome this barrier, recent approaches have turned towards on-device ML inference \cite{edgerec, google_on_device, on_device_xia, edge_neuron_surgeon}. Rather than send users' raw data to the cloud to perform model inference, users' raw data is directly processed on the users' devices, thereby eliminating the need for sensitive user data to be collected by centralized servers.

On-device ML inference is a promising solution to \rev{provide stronger privacy}, %stronger privacy regulations and policies \cite{edgerec, google_on_device, on_device_xia, edge_neuron_surgeon}, 
as it enables model inference without requiring clients to share private input features with the service provider. Unfortunately, a \rev{pure} on-device ML inference solution is impractical for many applications such as recommendation, as these applications often require access to an embedding table that is too large to store on device. For example, recommendation models access tables that often take gigabytes or even terabytes of memory \cite{deeprecsys, dlrm, dlrm_blog, embedding_codesign, din_ctr}. \rev{These embedding tables are accessed using user features that are important inputs to the recommendation model, and dropping them may negatively impact model quality.} Large embedding tables pose a dilemma: storing large embedding tables on device is impractical given device limitations while storing them in the cloud and directly accessing them in the clear  could leak private information.

To address this issue, we propose using private information retrieval (PIR) to privately query large embedding tables stored on servers. In this work, we consider distributed point function (DPF)-based PIR, in which private embedding lookups are performed by constructing and evaluating DPFs on two non-colluding servers (Figures \ref{fig:pir_on_device} and \ref{fig:dpf_pir_short}). A two-server DPF-PIR scheme is attractive as it is much more efficient in terms of computation and communication compared to single-server PIR schemes~\cite{pir_he, spiral}.
The two-server model is also widely used in the previous work on secure multi-party computation (MPC) for privacy-preserving machine learning \cite{falcon, ariann, cryptflow, crypten} or private analytics \cite{ipa,ipa_2}. 
% Kiwan: I think this should rather go to Sec 3 A
%In context of real-world applications, large companies seeking to deploy such a protocol would need collaboration from a neutral third party to act as the non-colluding second server. To satisfy this requirement, companies that form a consortium to establish privacy standards may act as each other's neutral third party. 

Despite their advantages, DPF-based PIR protocols still exhibit massive computational overhead \cite{dpf_1, dpf_2}, making them difficult to deploy in \rev{large-scale} applications \rev{that require high throughput}. The computational overhead stems from evaluating the DPFs on the servers, which entails executing a significant number of expensive cryptographic operations \cite{dpf_1, dpf_2}. For example, expanding a typical DPF for a table with one million entries requires performing at least one million AES-128 encryption operations. The cost is amplified during ML inference where a model may access multiple embedding entries \cite{deeprecsys, architectural_implication_recsys}.
The computation and communication requirements of DPF-based PIR make deploying it to real-world ML applications a considerable challenge. 

\begin{figure}
\centering

    \includegraphics[width=\linewidth]{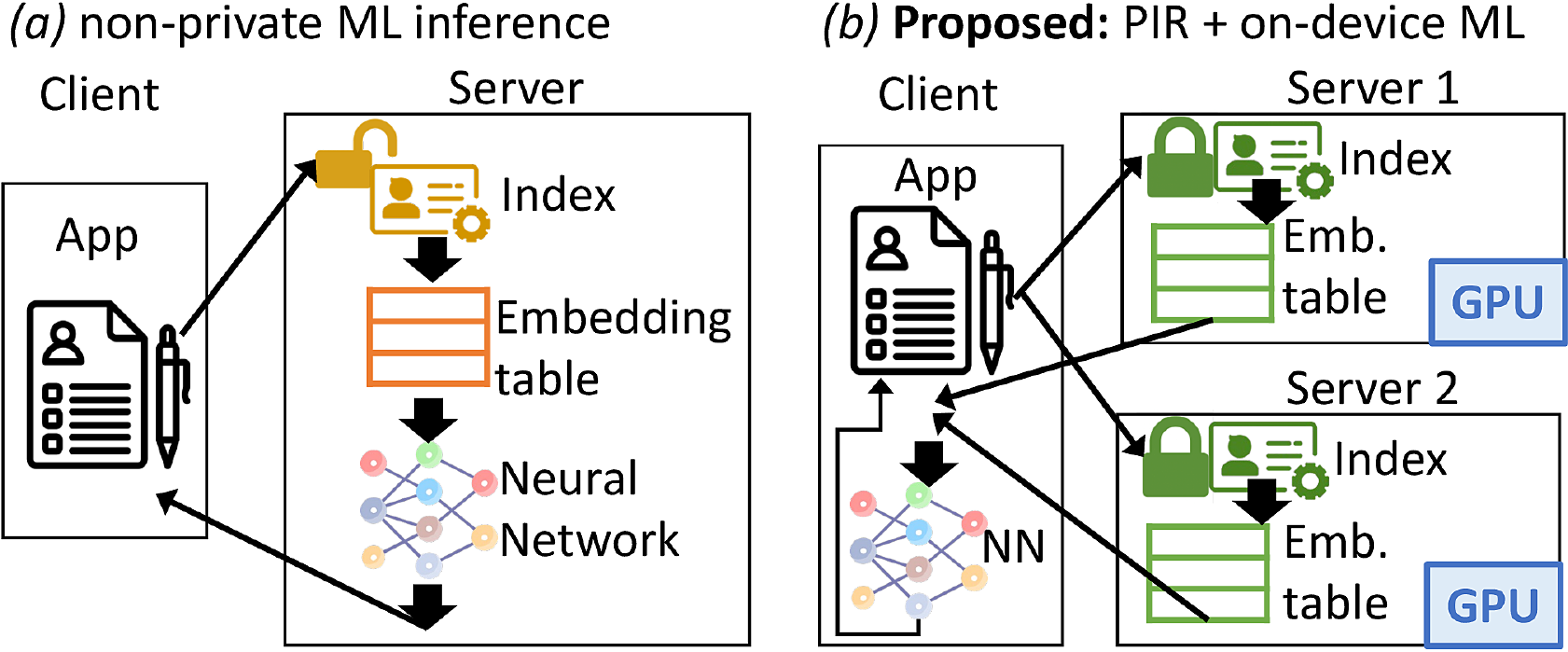}  
    \label{fig:tradinf}

    \caption{(a) The traditional non-private approach to ML inference, and (b) the proposed approach for private on-device ML inference. Using PIR, a CPU-based client privately obtains embeddings from two GPU-accelerated non-colluding servers; these embeddings are subsequently used as inputs to the client's on-device neural network. }
    \label{fig:pir_on_device}
    \label{fig:emb_lookup}
    \vspace{-.5em}
\end{figure}
\iffalse
\begin{figure}
\centering
\begin{subfigure}{.2\textwidth}
    \centering
    \hspace{-3em}
    \includegraphics[width=\linewidth]{figures/pir-on-device-1.pdf}  
    \caption{Traditional Non-private ML inference}
    \label{fig:tradinf}
\end{subfigure}
\quad
\begin{subfigure}{.2\textwidth}
    \centering
    \includegraphics[width=\linewidth]{figures/pir-on-device-2.pdf}  
    \caption{Proposed Approach: PIR + On-Device ML}
    \label{fig:proposedapproach}
\end{subfigure}
    \caption{Left: the traditional non-private approach to ML inference. Right: our proposed approach for private on-device ML inference. Using PIR, a CPU-based client privately obtains embeddings from two GPU-accelerated non-colluding servers; these embeddings are subsequently used as inputs to the client's on-device neural network.}
    \label{fig:pir_on_device}
    \vspace{-.5em}
\end{figure}
\fi

\subsection{Our Contributions}
We develop a system to efficiently and privately serve embeddings for on-device ML, \rev{with the primary focus on on-device recommendation models that require privately accessing large server-side embedding tables}. 
\rev{
Note that recommendation models represent an important application that account for a significant portion of the computational resources for ML in practice 
%(i.e: 80\% at Meta \cite{architectural_implication_recsys} and 25\% at Google cycles \cite{google_cycle}).} 
\cite{architectural_implication_recsys, google_cycle}.
While our work primarily targets private on-device recommendation, the proposed PIR system can also be applied to other on-device ML models that need private access to server-side embedding tables. 
%private information retrieval 
%(i.e., private inference of language models with tables stored on the server). 
}

Embedding accesses for on-device ML, \rev{particularly on-device recommendation}, have several unique properties and requirements compared to other applications that might use PIR: 1) embedding table entries are often short, between 64-1024 bytes, 2) multiple embedding table entries are often accessed together in a batch as part of a single model inference, and 3) throughput, latency, and model quality are all critical to an application's success. We leverage these properties to design a novel GPU acceleration scheme for efficiently performing PIR on GPUs, and, additionally, co-design PIR with the ML application to facilitate better trade-offs between model quality and system performance. 
%Our technical contribution is similar in its nature to how recent studies co-optimize algorithms and GPU implementations to significantly improve the performance of other cryptographic primitives such as fully-homomorphic encryption (FHE) \cite{spiral, floram, blinder, riposte}.   
Similar to other systems work in the PIR domain \cite{spiral, floram, blinder, riposte}, our contributions focus on performance improvements.. Our specific contributions are listed below.
% Our key contributions are discussed in more detail below.

%\textbf{Efficient GPU PIR}\\
\textbf{GPU-accelerated PIR}
We develop a set of novel optimizations to efficiently perform PIR on GPUs. Our optimizations enable high-throughput, low-latency DPF execution, allowing us to scale to tables with millions of entries. We observe that DPF evaluation is compute-bound due to their heavy cryptographic instruction mix, and leverage the fact that GPUs are especially well suited to perform these computationally heavy operations. Yet, performing PIR on a GPU requires exploiting multiple types of parallelism in PIR while carefully balancing computation, communication, and memory usage. Our GPU acceleration, over an optimized CPU baseline \cite{google_dpr}, obtains $>1,000 \times$ speedup over single-threaded CPU execution, and $>20 \times$ speedup over multi-core execution. 
To the best of our knowledge, 
this work represents the first to explore high-performance GPU implementations of DPFs.
%a high-quality GPU implementation of DPFs is not available to the public and our work is the first to explore and open-source such an implementation. 
We note that our GPU implementation accelerates the state-of-the-art DPF algorithm \cite{dpf_1}, which exhibits an optimal communication cost of $O(\log(n))$ and an optimal computation complexity of $O(n)$.
Beyond private embedding table accesses for ML, our GPU PIR can be used to accelerate any PIR applications such as checking compromised passwords. Our code is open sourced at \url{https://github.com/facebookresearch/GPU-DPF}.  

\textbf{ML + PIR Co-Optimization}
To further improve performance, we develop strategies utilizing application-specific data access patterns to co-optimize PIR with the ML application. Traditional batch PIR algorithms \cite{batch_codes, poly_codes, cuckoo}, which allow privately obtaining multiple entries together, may impact ML inference quality because they only retrieve entries probabilistically and may drop some queries. We co-design a new batch PIR algorithm for ML tasks to obtain a better trade-off between model quality and system performance. We comprehensively evaluate the resulting performance improvements and model quality of the new batch PIR scheme on applications including WikiText2 language model \cite{wikitext2}, Movielens recommendation \cite{movielens}, and Taobao recommendation \cite{taobao}. The results show that the proposed optimizations utilizing application-specific data access patterns can increase the ML inference throughput by up to $100 \times$ over a straightforward PIR system design on a multi-core CPU, while maintaining the model quality and limiting inference communication and latency within $300$ KB and $300$ ms, respectively.

%\section{\rev{Private On-Device ML Inference for Recommendation Systems}}
\section{Private On-Device ML Inference} % for Recommendation Systems}}

\subsection{\rev{Threat Model}}
%On-device ML inference is a solution to increasingly stringent privacy policies~\cite{on_device_xia, google_on_device}. Broadly, on-device inference performs ML computation on each user's device without sharing users' local private data with a service provider's server. 
%Private information may include users' click history, text messages, or search history, which is sensitive information users may be reluctant to share~\cite{apple_att}. 
The goal of private on-device inference is to perform ML inference using data on a user device without revealing them to a server owned by a service/cloud provider. 
\rev{In the context of recommendation systems, on-device inference can allow private user data only available on a client device to be used to provide more relevant recommendations, while ensuring that no private data leaves the device.
To reduce the burden on user devices, a server-side recommendation model can send a set of candidate recommendations based on less sensitive user features available on the server, then a smaller on-device model can more accurately rank the candidates leveraging private on-device user data without revealing them to the server.
In our study of a real-world model, we found that even a small (several MB) on-device MLP model can noticeably improve recommendation accuracy when combined with server-side embedding tables.
} 

We assume that the computation part of the ML model can run on the user device given the increasing trend of hardware accelerators and optimizations for client SoCs, but that \emph{embedding tables} of categorical/sparse features (described below) are too large to be placed on individual devices and hence are accessed remotely 
(Figure~\ref{fig:pir_on_device}). We further assume that only a very small fraction of the table is used per-inference. 
%% ED: the following doesn't really fit into the threat model section. It is more of a performance target. 
%We target a latency of 100-\rev{500}ms as these are typical latency requirements in company service level agreements \cite{architectural_implication_recsys}; for some applications, longer latency (around 1s) is accepted \cite{response_time}. 

\rev{
As the indices to embedding tables represent private categorical feature values, private on-device inference must ensure the confidentiality of table indices while allowing the use of server-side embedding tables. 
%The main As the confidentiality of user data is the main concern for privacy, our work focuses on confidentiality as in recent work~\cite{riposte, floram, crypten, cryptflow, blinder, apple_att, apple_att_repercusions}. 
For this purpose, we leverage private information retrieval (PIR) protocols under the honest-but-curious threat model. The user/client device and its software are trusted. 
While remote servers are untrusted, they are assumed to follow the protocol. 
The honest-but-curious threat model is widely used in previous private inference work \cite{crypten, cryptflow, cryptgpu, blinder, riposte, delphi}. 
The model may be extended to a malicious setting by using PIR protocols that protect against a malicious server deviating from the protocol and produce wrong answers (e.g. authentication for PIR \cite{auth_pir}).
We also note that incorrect PIR responses only lead to non-optimal suggestions in recommendation models; selective failure attacks \cite{selective_failure} are difficult to perform because failures are not visible to attackers. 
}  

Like previous work on privacy preserving ML and analytics using multi-party computation (MPC) \cite{blinder, riposte, floram, falcon, ariann, cryptflow, crypten,ipa,ipa_2}, we further \rev{assume} a two-server model where the two servers are non-colluding. 
This two-server setup can be practically realised by having two different cloud vendors host and manage the two servers or having another industry actor host the second server. Forming such a privacy consortium among companies is emerging in industry~\cite{mpc_alliance}. See Section~\ref{sec:related} for further discussions.
%A possible alternative is to use new services that can attest the code running on them, such as Microsoft Azure's Confidential Computing framework~\cite{ACC}, to act as a trusted non-colluding party. See Section~\ref{sec:related} for further discussions.} 
%Another practical method is to have the second server be realized by vendors like AWS or Microsoft Azure. New services such as Microsoft Azure's confidential computing framework provide services similar to what our approach require\cite{ACC}. See Section~\ref{sec:related} for further discussions.} 

% This work only considers the confidentiality of the user data and does not consider the integrity/correctness of the inference result. 
% Figure~\ref{fig:pir_on_device} compares the traditional cloud-based ML services and the proposed on-device ML.

\begin{table}%[tb!]
\centering
\caption{Embedding table sizes for popular public datasets and models spanning across language and recommendation.
%Table sizes, in context of on-device ML inference, range from reasonably small to impractically large.
}
\label{tab:emb_size}
{\footnotesize %scriptsize
\begin{tabular}{|c|c|c|c|}
\hline
\textbf{Application}                                                                      & \textbf{\begin{tabular}[c]{@{}c@{}}\# of \\ Embedding \\ Entries\end{tabular}} & \textbf{Entry Size} & \textbf{\begin{tabular}[c]{@{}c@{}}Embedding \\ Table Size\end{tabular}} \\ \hline
\textbf{\begin{tabular}[c]{@{}c@{}}Criteo 1 TB\\ Rec.\end{tabular}}             & \textgreater{}4,000,000,000                                                        & $\sim$128B          & \textgreater 476 GB                                                      \\ \hline
\textbf{\begin{tabular}[c]{@{}c@{}}Criteo \\ Rec.\end{tabular}}                 & $\sim$45,000,000                                                                   & $\sim$128B          & $\sim$5 GB                                                               \\ \hline
\textbf{\begin{tabular}[c]{@{}c@{}}FastText Emb.\\ (Language Model)\end{tabular}} & $\sim$2,000,000                                                                    & $\sim$1024B         & \textgreater 1.9 GB                                                      \\ \hline
\textbf{\begin{tabular}[c]{@{}c@{}}Taobao \\ Rec.\end{tabular}}                 & $\sim$900,000                                                                      & $\sim$128B          & $\sim$109 MB                                                             \\ \hline
\textbf{\begin{tabular}[c]{@{}c@{}}WikiText2\\ (Language Model)\end{tabular}}             & $\sim$131,000                                                                      & $\sim$512B          & $\sim$64 MB                                                              \\ \hline
\textbf{\begin{tabular}[c]{@{}c@{}}Movielens-20M\\ Rec.\end{tabular}}         & $\sim$27,000                                                                       & $\sim$128B          & $\sim$3 MB                                                               \\ \hline
%\hhline{|=|=|=|=|}
%\textbf{\begin{tabular}[c]{@{}c@{}}Real-World Feature-1\end{tabular}} & 7,614,589 & 144B & 1.02GB \\ \hline
%\textbf{\begin{tabular}[c]{@{}c@{}}Real-World Feature-2\end{tabular}} & 20,000,000 & 144B & 2.68GB \\ \hline
%\textbf{\begin{tabular}[c]{@{}c@{}}Real-World Feature-3\end{tabular}} & 20,000,000 & 144B & 2.68GB \\ \hline
%\textbf{\begin{tabular}[c]{@{}c@{}}Real-World Feature-4\end{tabular}} & 2,989,943 & 144B & 410MB \\ \hline
%\textbf{\begin{tabular}[c]{@{}c@{}}Real-World Feature-5\end{tabular}} & 20,000,000 & 144B & 2.68GB  \\ \hline

\end{tabular}
}

\vspace{-10pt}
\end{table}

\subsection{Key Challenge: Large Embedding Tables}
%\mrhu{The paper as-is right now doesn't have a background section that explains what an embedding table is (i.e., given an index ID as input, you get an output vector). Adding a paragraph with a simple figure (Udit should have many examples on this?) explaining the overall embedding lookup process will help the reviewers better appreciate where PIR fits in.}

Unfortunately, the embedding tables in machine learning models, 
\rev{especially for recommendation models}, %that \rev{recommendation systems} employ 
are often too large for individual devices \cite{deeprecsys, dlrm, dlrm_blog, embedding_codesign, din_ctr}, making a pure on-device inference solution impractical. 
An embedding table is a large table that maps categorical features into dense vectors that encode semantic information. For example, categorical (sparse) features may include a user's click or search history. The value of a categorical feature is used as an index to an embedding table where each row of the table holds the vector corresponding to that categorical feature value (Figure~\ref{fig:emb_lookup}). Embedding tables can have as many rows as the number of possible values in the categorical feature space so their size can grow quickly. 
%\rev{Below, we detail the role that embedding tables play in recommendation system, and, as a minor example, in language models, to demonstrate their wide application in both recommendation systems and other ML applications.}

\textbf{Recommendation models} %, \rev{which are our primary target use-case due to their reliance on large server-side tables,} 
%Recommendation models, \rev{which represent the primary target for our system},  
use several user and product input features to predict whether a user is likely to interact (e.g., click or purchase) with the product~\cite{dlrm, din_ctr}. These models may use user data such as the list of products the user recently purchased~\cite{din_ctr}. As the number of products can be on the order of millions, the corresponding embedding table can reach several GB to TB in size~\cite{deeprecsys, embedding_codesign, edgerec}.
%Even though some systems may use a hash to reduce the size of an embedding table by sharing embedding table entries among multiple feature values, the embedding tables in real-world recommendation systems are still quite large.
Compressing the table is difficult for many real-world models, as it leads to significant accuracy drop~\cite{baidu_hierarchical}.
\rev{Recommendation models represent our primary target use case given their reliance on large server-side tables.} 

%\subsection{Example: Language Models}
\textbf{Language models} 
%\rev{
%albeit a more minor use-case due to newer language models employing smaller embedding tables, are 
\rev{are another potential example of an ML application that may require access to server-side embedding tables.} Language models empower applications such as next-word prediction, language translation, and speech recognition. Language models map words into a latent embedding space using word embedding tables~\cite{wikitext2}. As there may be hundreds of thousands of different words, with each embedding vector being hundreds of bytes long, it quickly becomes impractical to store the entire word embedding table on-device, especially for natural language translation models supporting multiple languages \cite{word_embedding_advantage, nllb}. 
Although there are alternative techniques to compress \rev{the embeddings} (e.g., character embeddings, sentence level representations, etc.),
word embeddings are considered to be more efficient to train in a regime with less training data \cite{word_embedding_advantage}. 
\rev{
We discuss the language model as a potential example in our study to show that our system can be adopted for multiple types of on-device models that need large server-side embedding tables.
However, we note that on-device inference for language models is limited to smaller language models that can run on a client device. Private inference for large language models need additional computation beyond embedding table accesses to be securely offloaded to cloud servers.
Also, the embedding tables for language models are typically much smaller compared to the tables for recommendation models.
}
%\rev{We highlight that these advanced techniques reduce the size of the tables for language models, and hence reduce their importance as they may be able to be stored on-device. Hence our work primarily focuses on recommendation systems which involve large embedding tables too large to store on-device. However, the language model example serves to demonstrate that various other ML applications may involve server-side embedding tables, for which our work would be immediately applicable.}

\iffalse
\begin{figure}
    \centering
    \includegraphics[width=.5\columnwidth]{figures/pir-embedding-access.pdf}
    \caption{Embedding tables map indices -- numerical ids that represent information like user's text messages or search history -- into vectors of features that are inputs to a neural network model.}
    \label{fig:emb_lookup}
    \vspace{-.5em}
\end{figure}
\fi

%
Table~\ref{tab:emb_size} summarizes the size of the embedding tables of some popular datasets/models. The size ranges from several MBs to hundreds of GBs. 
On the other hand, the mobile app size is on average 34MB, and seldom exceeds 200MB even in extreme cases~\cite{app_file_size}. Embedding tables, especially for \rev{recommendation models}, can easily exceed this range, % (Table \ref{tab:emb_size}),
which makes deploying them on-device impractical~\cite{edgerec}.  

%Unfortunately, many ML models require access to a large table of embeddings -- vector representations of items -- that are too large to store on-device. For instance, in recommendation models, user's product click history may be an input feature. This product click history is typically represented as a list of numerical ids that index into a table of features that need to be retrieved by a device. Performing model inference entails looking up and retrieving these embeddings to be used as input to the model. This embedding table, which may span millions of entries, one for each possible product in existence, is impractical to store on-device due to its size \cite{embedding_codesign, edgerec}. Another example is next word prediction, where user's words are directly mapped to numerical features, which are in turn inputted into the ML model~\cite{wikitext2}. As there may be hundreds of thousands of words, with each embedding entry being hundreds of bytes long, it quickly becomes impractical to store the entire embedding table on-device, especially in the context of multiple languages \cite{word_embedding_advantage, nllb}. Table TODO shows the extent of the size of some of these embedding tables that are used in practical applications. Hence, the size of embedding tables pose a significant challenge to a purely on-device solution to private ML inference.

% NLLB-200: https://ai.facebook.com/blog/nllb-200-high-quality-machine-translation/

\subsection{Example: Real-World Recommendation Model}

%We also include statistics on the top-5 most important tables (as evaluated using an internal feature-importance metric) for a industrial real-world on-device ML application. In this application multiple \emph{large} tables are queried for a single inference; these tables are too large to practically store all on the user's device and hence motivates PIR for private table accesses. 

%An example of the top-5 tables queried in a real-world on-device ML application is listed in Table \ref{tab:emb_size} (with "Feature-x" in their names); in this application, multiple large tables too large to practically store on-device are queried per inference. The large memory requirements of embedding tables prevent them from being stored on-device, and hosting them on the server may be the only practical solution.

As a concrete use case for private on-device ML inference with sparse features, we studied a real-world recommendation model where some of its input (user) features can only to be used on a client device for strong privacy protection. 
For this model, such ``device-only'' sparse features represent 7 out of top 25 features when the input features are ranked by their feature importance score\footnote{This score measures the change in the accuracy when a particular feature is changed to a random value.}.
Removing the device-only features significantly degrade the model's utility (accuracy), and a small (several MB) on-device model can provide good accuracy if the embedding tables can be accessed privately.  

\begin{table}%[tb!]
\centering
\caption{The embedding tables for a real-world recommendation model, showing the number of entries, the table size, and the average number of entries accessed per inference. The numbers are shown for the top 5 device-only sparse features with highest importance.}
\label{tab:industry_dlrm}
{\small %scriptsize
\begin{tabular}{c|c|c}
\hline
\# Entries & \begin{tabular}[c]{@{}c@{}}Avg Queries \\ Per Inference\end{tabular}  & \begin{tabular}[c]{@{}c@{}}Table Size \\ (\# of entries * 144B) \end{tabular}\\ \hline
7,614,589 & 13.9 & 1.02GB \\ \hline
20,000,000 & 47.3 & 2.68GB \\ \hline
20,000,000 & 25.7 & 2.68GB \\ \hline
2,989,943 & 3.2 & 400MB \\ \hline
20,000,000 & 14.9 & 2.68GB \\ \hline

\end{tabular}}
\vspace{-5pt}
\end{table}

Table~\ref{tab:industry_dlrm} shows the embedding table size and the number of accesses per inference for the top 5 sparse features that are only accessible on-device. Similar to the public datasets, the embedding tables are too large to be sent and stored on a client device, and each table entry is relatively small (144 bytes) -- on average only at most 1-10KB of entries are fetched from the table for each inference.

Our study also found that the user features change relatively slowly; the sparse user features mostly stay the same for two consecutive recommendations for one user. If a client device keeps recently fetched embedding table entries, only 2.44\% of sparse features are new and need to access embedding tables on a server. Even though Table~\ref{tab:industry_dlrm} shows that several tens of embedding table entries are used for each inference, the temporal locality means that only a few new entries need to be read from the server.

\subsection{Our Approach: On-Device ML Inference with PIR}

To enable private on-device ML applications that require access to large embedding tables, we propose using private information retrieval (PIR)~\cite{pir,floram}. 
PIR allows a user to query a table without revealing which index was accessed to the table holder, i.e., the server that hosts the embedding table.
We propose to keep large embedding tables on the cloud servers, and use PIR to query the table upon an embedding table access by a client's device (Figure \ref{fig:pir_on_device}).

We use a PIR protocol based on a distributed point function (DPF) \cite{dpf_1, dpf_2}, which protects accesses using two non-colluding servers. We choose PIR rather than oblivious RAM (ORAM) \cite{oram_orig,path-oram,wang2015circuit, aegis,intel-sgx,amd-sev,arm-trustzone, fletcher2015freecursive,ren2013design,wang2017cooperative,wang2018d,hardware_oram,ps_oram, laoram, spatial_oram, xiong2022secndp}, another popular cryptographic technique to hide an access pattern to memory, because ORAM is designed to protect accesses from a single entity. In the on-device ML scenario, multiple users simultaneously send query requests. DPF-based PIR methods are more efficient in terms of communication and computation compared to single-server PIR schemes that employ homomorphic encryption \cite{spiral, pir_preprocess, pir_he, pir_in_storage}. %; see related works for a comparison \cite{pir_in_storage}. 
A key challenge in employing DPF-based PIR is its high computational intensity due to heavy cryptographic operations. In the following section, we describe how the DPF-based PIR can be efficiently accelerated on GPUs. % and discuss its inherent parallelism that makes them amenable for GPU acceleration.

%Parallelizing these workloads on massive GPUs is a natural solution; however, due to their tree-like computation structure, parallelizing PIR on GPUs is tricky. In the next section we briefly describe PIR computation workloads, and introduce our novel algorithm for accelerating them on GPUs.

\section{Accelerating PIR using GPUs}

Algorithms for PIR exhibit significant overhead due to their heavy cryptographic operations and cannot be immediately adopted for private on-device inference. Below, we 1) briefly introduce PIR and DPF, 2) analyze their characteristics to understand how GPUs may accelerate them, and 3) describe our optimizations for GPU acceleration.

\subsection{Fundamentals of PIR and DPF}
\label{sec:dpf}
\begin{figure}
    \centering
    \includegraphics[width=.7\columnwidth]{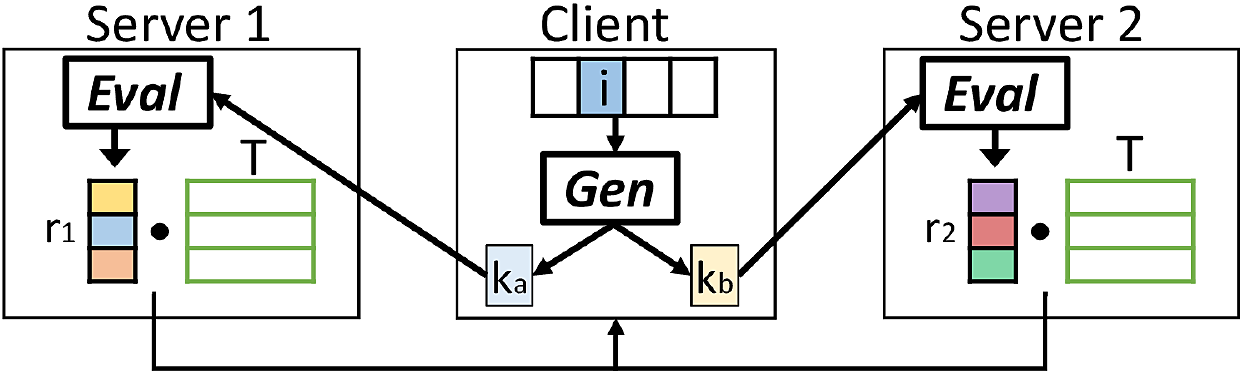}
    \caption{DPF based PIR scheme. The client computes $Gen$ to obtain two keys ($k_a$, $k_b$) that represent a secret index and sends them to the servers. The servers individually compute $Eval$ to obtain secret shares of the answer, from which the client can later retrieve the desired embedding. $Eval$ is computationally expensive and is our main target for acceleration.}.
    \label{fig:dpf_pir_short}
    \vspace{-5pt}
\end{figure}

Private information retrieval (PIR) based on distributed point functions (DPF) allows a user to access an index in a table, shared across two non-colluding servers, without leaking the index to the table holders. 
%Broadly, we consider private information retrieval (PIR) techniques based on distributed point functions (DPF).
In DPF-PIR, the client sends a key that represents the index it wants to privately query. The server, upon receiving the key, performs expensive cryptographic operations to service the user's query (Figure~\ref{fig:dpf_pir_short}).
%Accelerating DPF-based PIR using GPUs is critical to obtaining an efficient system for PIR.

\noindent \textbf{Naive PIR}
Assume a client $C$ seeks to privately access entry $T[i] \in \mathbb{F}_p^{D}$ from a table $T \in \mathbb{F}_p^{L \times D}$ that is duplicated across two non-colluding servers, $S_1$ and $S_2$. Here, $L$ is the number of entries in the table,  $D$ is the vector length of each entry, and $\mathbb{F}_p$ is an integer field with modulus $p$. A simple but highly inefficient approach is for the client $C$ to generate and send a random vector $r_1 \in \mathbb{F}_p^{1 \times L}$ and a second vector $r_2 \in \mathbb{F}_p^{1 \times L}$ to $S_1$ and $S_2$, such that they add up to a one-hot indicator vector $I(i)$ whose entries are all 0's except at the $i^{th}$ position where it is 1 ($r_1 + r_2 = I(i)$). Upon receiving the vectors, the servers individually compute and return $r_1 \times T$ and $r_2 \times T$ to the client, from which the client can retrieve $T \times (r_1 + r_2) = T \times I(i) = T[i]$. Information theoretic privacy is ensured as $r_1$ and $r_2$ are {\em secret shares} of the indicator vector that do not leak any information about $i$ individually~\cite{secret_share}. This simple approach incurs large communication overhead because the size of $r_1$ and $r_2$ is proportional to the size of table $T$, making the communication overhead $O(L)$.

\noindent \textbf{DPF-PIR}
The generalization of the approach described above is a cryptographic primitive known as a \emph{distributed point function} (DPF). DPF is an algorithmic construct that allows a client to \emph{generate} two compact keys $k_a$, $k_b$, such that when the keys are \emph{expanded} across a set of indices, they yield secret shares of the indicator vector $I(i)$. 

%At a high level, the expansion involves traversing through a binary tree from the root and calling a series of \emph{pseudorandom function}s (PRFs; e.g., AES-128) with the output of the parent node (Figure~\ref{fig:pir_figs}).
%
Formally, a DPF consists of two algorithms,
\begin{itemize}
    \item $Gen(1^\lambda, i \in {0 .. L-1}) \rightarrow (k_a, k_b)$, which takes security parameter $\lambda$ and input $i$, and generates two keys $k_a$, $k_b$.
    \item $Eval(k, j) \rightarrow \mathbb{F}_p$, which takes a key $k$ and an evaluation index $j$ and outputs a field element.
\end{itemize}
such that, $Eval(k_a, j) + Eval(k_b, j) =  \begin{cases} 
      1 & j=i \\      
      0 & j \neq i\\
   \end{cases}.
$

\begin{figure}[tb!]
    \centering
    \includegraphics[width=.7\linewidth]{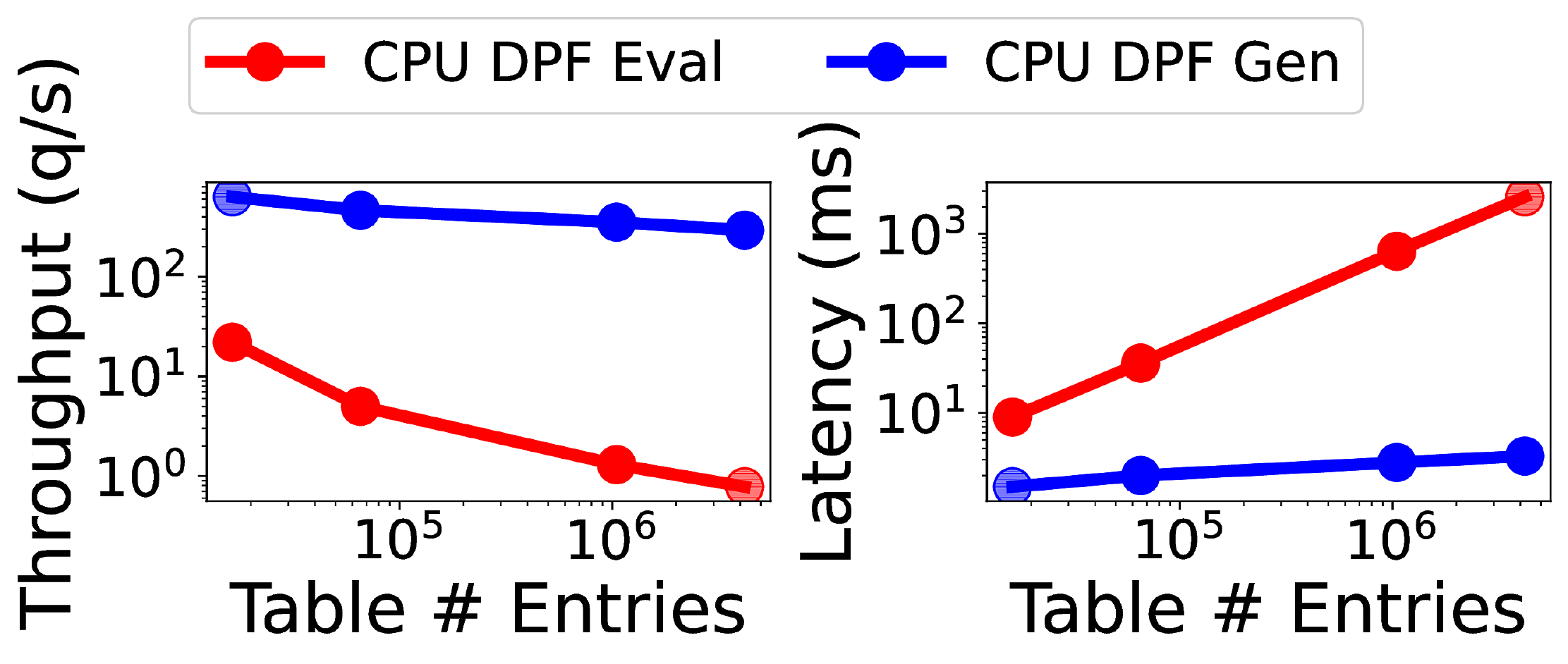}
    \caption{$Gen$ vs $Eval$ performance. $Gen$ is highly efficient and is not our target for optimization.}
    \label{fig:keygen}
    \vspace{-5pt}
\end{figure}

$Gen$ is a key generation process where a client encrypts the index it wishes to query into two keys $k_a$ and $k_b$, which are respectively sent to the two non-colluding servers.
$Gen$ is relatively lightweight compared to $Eval$ ($O(log(L)$ computation)~\cite{dpf_1, dpf_2}, and can be quickly computed even on resource-constrained client devices as shown in Figure \ref{fig:keygen}.

$Eval$ is the key evaluation process that is performed on the servers. 
Upon receiving $k_a$ or $k_b$, the servers respectively compute $T \times Eval(k_a, \{0 \ldots, L-1\})$ and $T \times Eval(k_b, \{0 \ldots, L-1\})$ and return the result, from which the client can obtain $T \times (Eval(k_a, \{0 \ldots, L-1\}) +  Eval(k_b, \{0 \ldots, L-1\})) = T \times I(i) = T[i]$. $Eval$ requires at least $O(L)$ computation \cite{dpf_1, dpf_2} and is the major bottleneck (see Figure \ref{fig:keygen}). Our work focuses on accelerating the $Eval$ function.
Figure \ref{fig:dpf_pir_short} depicts the overall DPF-PIR scheme.

A DPF should be computationally secure, meaning that given just one of the keys and no other information, it should be difficult to recover the client-queried index $i$ without doing computation proportional to $O(2^\lambda)$.
%Given a DPF, client $C$ can generate keys $k_a$, $k_b$ using $Gen$ and send the keys to $S_1$ and $S_2$, respectively. The two servers, upon receiving $k_a$ and $k_b$, compute $T \times Eval(k_a, \{0, 1, \ldots, L\})$ and $T \times Eval(k_b, \{0, 1, \ldots, L\})$ and return the result, from which the client can obtain $T \times (Eval(k_a, \{0, 1, \ldots, L\}) +  Eval(k_b, \{0, 1, \ldots, L\})) = T \times I(i) = T[i]$. Figure \ref{fig:dpf_pir_short} depicts the overall DPF-PIR scheme.
%
There are many different implementations of DPFs, each with a different computation/communication trade-off. We consider the DPF construct described in~\cite{dpf_1}, which provides optimal asymptotic communication complexity of $O(\lambda  \log(L))$ and optimal evaluation computation complexity of $O(\lambda  L)$.
%In this DPF algorithm, key $k$ \lhl{consists of } two codewords $\{C_0 \in \mathbb{F}_{2^\lambda}^{2 \times \log(L)}, C_1 \in \mathbb{F}_{2^\lambda}^{2 \times \log(L)}\}$. DPF evaluation \lhl{involves computing the value of the leaves of the computation tree specified by the following equation:}

%$$
%Eval(k, i) = Expand(C_0, C_1, d=log(L), i)
%$$

%\lhl{$d$ is the current height of the node that is being expanded (0 being the root node, $log(L)$ specifying the leaves), and $i$ is the index of the node at that height (0 being the left-most node). Then, DPF evaluation is specified by the following recurrence:}

%$$
%Expand(C_0, C_1, d, i) = PRF_{P}(i \mbox{ mod } 2) + C_{P \mbox{ mod } 2}[i \mbox{ mod } 2, d]\\
%$$
%$$
%P = Expand(C_0, C_1, d-1, \lfloor \frac{i}{2} \rfloor)
%$$

%\lhl{$P$ is a recursive call that evaluates the node's parent's value and is a 128-bit value, and is used as both a key to the $PRF$ (pseudorandom function, i.e: AES-128), and to select which codeword of $k$ (either $C_{0}$ or $C_{1}$) to use. The index $i$ and depth $d$ of the node are analogously used as the input to the PRF and to index into the selected codeword. Evaluating this recurrence is equivalent to expanding a binary-tree (Figure \ref{fig:pir_dpf_recurrence}).}

\begin{figure}
    \centering
    \includegraphics[width=.8\columnwidth]{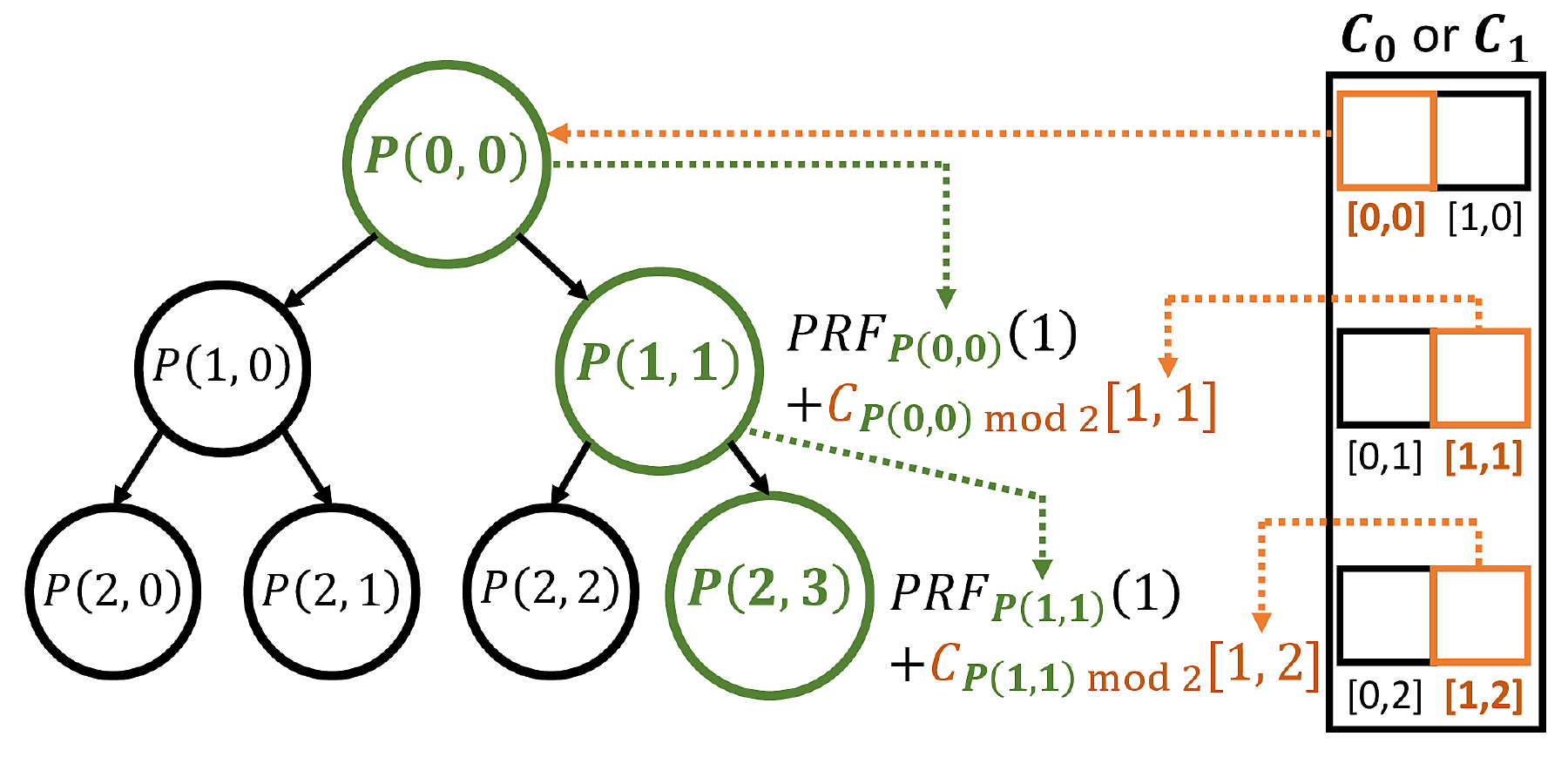}
    \caption{Example of the DPF computation using tree expansion. DPF expansion involves computing the leaves of a binary computation tree which evaluate to a secret-share of a one-hot vector. Computing each node requires evaluating its parent node which involves calling a PRF and adding to it a a codeword value indexed by the height and parity of the node.}
    \label{fig:pir_figs}
    \vspace{-5pt}
\end{figure}

In this DPF algorithm, the evaluation of DPF involves expanding a GGM-style \cite{ggm} computation tree. Keys $k_a$ and $k_b$ each consists of two two-dimensional codewords, $\{C_0 \in \mathbb{F}_{2^\lambda}^{2 \times (\log(L)+1)}, C_1 \in \mathbb{F}_{2^\lambda}^{2 \times (\log(L)+1)}\}$. The server uses the codewords and expand them into a tree (Figure~\ref{fig:pir_figs}) to get the secret shares of the indicator vector, using a recursively-defined helper function $P$:

\begin{equation}
Eval(k, j) = P(d=log(L), j)
\label{eq:p1}
\end{equation}
\begin{equation}
P(0, 0) = C_{0}[0, 0]
\end{equation}
\begin{equation}
\begin{aligned}
P(d, j) &= PRF_{P(d-1, \lfloor \frac{j}{2} \rfloor)}(j \mbox{ mod } 2) \\&+ C_{P(d-1, \lfloor \frac{j}{2} \rfloor) \mbox{ mod } 2}[j \mbox{ mod } 2, d]
\end{aligned}
\label{eq:p3}
\end{equation}

Here, $d$ is the depth of the node (0 for the root, $log(L)$ for the leaves), $j$ is the index of the node within each depth (0 being  leftmost), and $PRF_{s}(x)$ is a \emph{pseudorandom function} that encrypts a message $x$ with an encryption key $s$, such as AES-128.

Figure~\ref{fig:pir_figs} illustrates how $Eval$ works with an example. Assume the client wants to query a table of $L=4$. The client generates and sends a key to each server, where each key consists of two 2$\times$3 codewords, $C_0$ and $C_1$. Using the keys, the server must calculate $Eval(k, 0)..Eval(k, 3)$ and multiply them to the table. To calculate, \emph{e.g.}, $Eval(k, 3)$ (which is $P(2, 3)$ from Equation~\ref{eq:p1}), the server needs to calculate $P(1, 1)$, calculating which in turns requires $P(0, 0)$ (Euqation~\ref{eq:p3}). The calculation can be seen as an evaluation of each node in a binary tree from the root to the leaf; a child node is computed using the result from the parent node and $C_0$, $C_1$.

%\lhl{As an example, assume the client wants to lookup a table of size 4. }

Evaluating a single node requires a single $PRF$ call and an addition, requiring $O(\lambda L)$ computation for the entire tree. Communication overhead is proportional to the size of the keys, resulting in $O(\lambda \log(L))$ total communication. In practice, $\lambda$ is typically a 128-bit field integer to ensure sufficient computational security.
%A figure depicting the DPF computation tree and its recurrence relation is shown in Figure \ref{fig:pir_dpf_recurrence}.
After computing all the leaf nodes of the tree, the output is a vector of $\lambda$-bit (128-bit) field values; the final secret shares of the entry are obtained by performing an integer dot product between the computed 128-bit field values and the table. \rev{Note that tables with \emph{arbitrary} sized entries (i.e: much greater than 128-bits) may be supported with no additional DPF evaluation, as we can view these large-entried tables as a 2-D matrix, with the large entries subdivided into groups of 128-bit values; we may then perform a matrix-vector-multiplication with the prior DPF output to obtain secret shares of the table lookup. This works as performing a matrix-vector-multiplication between the DPF vector and the 2-D table selects the entire set of entries that corresponds to the selected index. In practice, the dot products for multiple queries to a single table are batched together as a single matrix-matrix multiplication to enhance performance.} We refer to \cite{dpf_1} for details on key generation.

%\begin{figure}
    %\centering
    %\begin{subfigure}{.23\textwidth}
    %\centering
    %\includegraphics[width=.85\columnwidth]{figures/pir-dpf.pdf}
    %\caption{DPF Computation structure. \label{fig:pir_dpf_recurrence}}
    %\end{subfigure}
    %\begin{subfigure}{.23\textwidth}
    %\centering
    %\includegraphics[width=.85\columnwidth]{figures/pir-dpf-exposed-parallelism.pdf}
    %\caption{Parallelizing DPFs. %\label{fig:dpf_expose_par}}
%    \end{subfigure}
%   \label{fig:dpfs}
%   \caption{Computation structure of DPFs (left) and parallelism opportunities (right).}
%    \vspace{-10pt}
%\end{figure}

\subsection{Accelerating PIR with GPU}

\begin{figure}[t]
\centering
\includegraphics[width=1\linewidth]{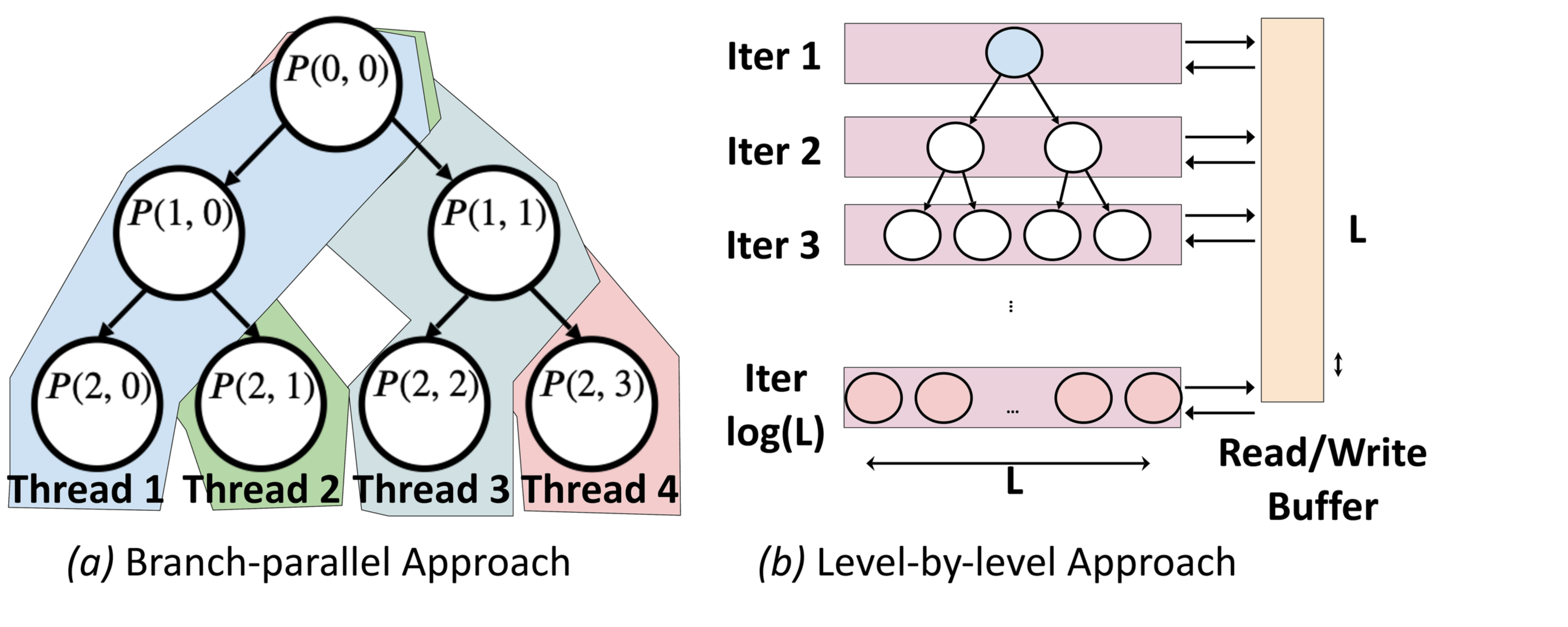}  
\iffalse
\begin{subfigure}{.22\textwidth}
    \centering
    \includegraphics[width=.9\linewidth]{figures/New_PIR_Naive_Strat.pdf}  
    \caption{Branch-parallel Approach}
    \label{fig:naive_dpf_par}
\end{subfigure}
\begin{subfigure}{.22\textwidth}
    \centering
    \includegraphics[width=.9\linewidth]{figures/pir-opt-2.pdf}  
    \caption{Level-by-level Approach}
    \label{fig:dpf_redundancy_plot}
\end{subfigure}
\fi
\caption{Two naive approaches for parallelizing DPF computation.}
\label{fig:naive_par}
\vspace{-5pt}
\end{figure}

\subsubsection{Starting Point: Batched DPF Execution}
We begin by observing that parallelism in DPF computation can be exposed in two dimensions: 1) parallelizing the evaluation of a \emph{single} DPF; and 2) evaluating \emph{multiple} DPFs in parallel. The latter, evaluating multiple DPFs in parallel, is understood as standard \emph{batched execution} and is an implicit starting point inherent to our proposed optimizations. At the GPU level, parallelizing the evaluation of a single DPF is done via thread-level parallelism, and batched-execution is performed by evaluating multiple  DPFs on multiple blocks via block-level parallelism. Under this framework, approaches falling under the two categories can be applied jointly with minimal interaction, and hence, unless otherwise noted, batched-execution with batch-size $B$ is assumed in all subsequent parallelization approaches. 
While batching itself is not a novel component of our proposed approach, batching is indeed important for high utilization of GPU resources (Figure \ref{fig:util_batch}). We also found that the batch size needs to be carefully selected based on the size of the table and the DPF paralleization strategy to balance latency, throughput, and memory requirement.

\subsubsection{Tradeoffs between Branch-parallel and Level-by-level DPF Parallelization Approaches}
\label{sec:bp_and_lvl}
% +E
Two naive approaches to parallelizing the execution of individual DPFs are branch-parallel and level-by-level approaches, shown in Figure \ref{fig:naive_par}. A branch-parallel approach has each thread independently compute one branch/leaf (or a subset of branches/leaves) of the DPF, while a level-by-level parallelization approach has each thread evaluate the nodes of a single level of the DPF tree in parallel, writing outputs to global memory to be used for computing the next level.

Unfortunately, these two naive parallelization approaches suffer from a major tradeoff between computational redundancy and memory usage, making neither truly efficient nor scalable. A branch-parallel approach suffers from \emph{computational redundancy}. As computing each leaf node requires evaluating all nodes up to the root, each thread in branch-parallel execution re-computes intermediate nodes unnecessarily, as shown in Figure~\ref{fig:naive_par}a.
As a result, the overall amount of work becomes $O(L \cdot log (L))$, instead of the optimal $O(L)$. 

The level-by-level parallelization approach eliminates this computational redundancy by storing and reusing intermediate node outputs. However, this approach suffers from \emph{memory overhead} as storing intermediate results consumes significant amount of memory when the batch size and the table size is large ($O(BL)$ for a batch size $B$). %space which, on large tables run into memory limitation issues and necessitates reducing $B$ batch-size, which reduces parallelism.
Hence, there is a fundamental tradeoff between these two approaches in balancing computation and memory usage. Figure~\ref{fig:redundant} shows that the branch-parallel approach suffers from high number of PRF calls, while the level-by-level approach suffers from high peak memory usage.

\subsubsection{Memory-bounded Tree Traversal}% for Efficient DPF Parallelization} % +E +M

\begin{figure}[t]
\centering
 \includegraphics[width=\linewidth]{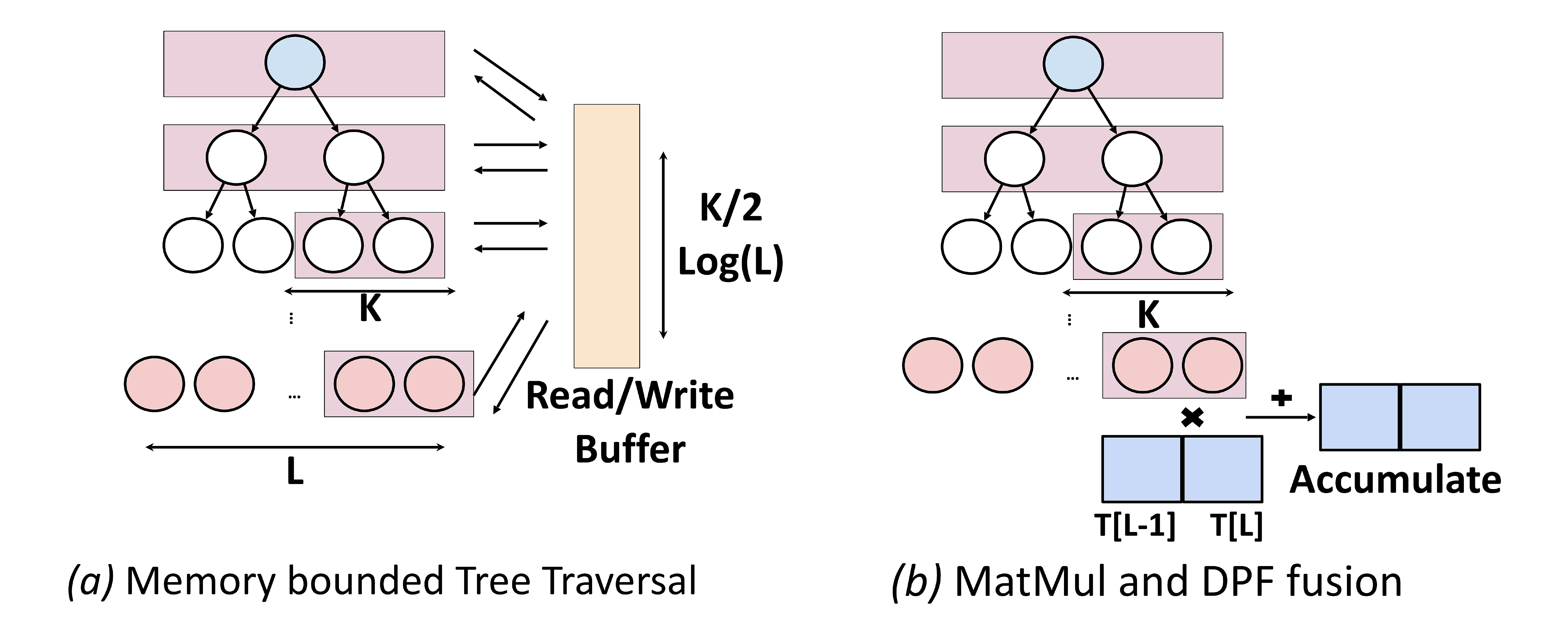}  
\iffalse
\begin{subfigure}{.22\textwidth}
    \centering
    \includegraphics[width=.9\linewidth]{figures/pir-opt-3.pdf}
    \caption{Memory-bounded Tree Traversal}
    \label{fig:dfs}
\end{subfigure}
\begin{subfigure}{.22\textwidth}
    \centering
    \includegraphics[width=.9\linewidth]{figures/pir-opt-4.pdf}  
    \caption{Matrix-multiplication and DPF Operator Fusion}
    \label{fig:fuse}
\end{subfigure}
\fi
\vspace{-15pt}
\caption{Memory-bounded tree traversal and operator fusion for reducing memory overhead.}
\label{fig:mem_bound_tree_traversal_plus_fusion}
\end{figure}

The tradeoff between computation and memory usage in Section~\ref{sec:bp_and_lvl} motivates a different parallelization strategy. \rev{We emphasize that memory usage is a critical factor in accelerating DPFs on GPUs, as memory limitations bound the effective batch size that may be used; consequently, reducing memory usage allows for the use of larger batch sizes which significantly increases throughput. In other words, reducing memory usage while ensuring efficient parallel execution is the key to efficient DPF acceleration on a GPU.} To this end, we develop \emph{Memory-bounded tree traversal} (Figure \ref{fig:mem_bound_tree_traversal_plus_fusion}a), a parallelization scheme that is: 1) optimal in terms of computation ($O(L)$ work); and 2) exhibits memory overhead that scales \emph{logarithmically} with the size of the table, instead of linearly as in the level-by-level approach.

\begin{figure}[t]
\centering
%\begin{subfigure}{.42\textwidth}
    \centering
    \includegraphics[width=.7\linewidth]{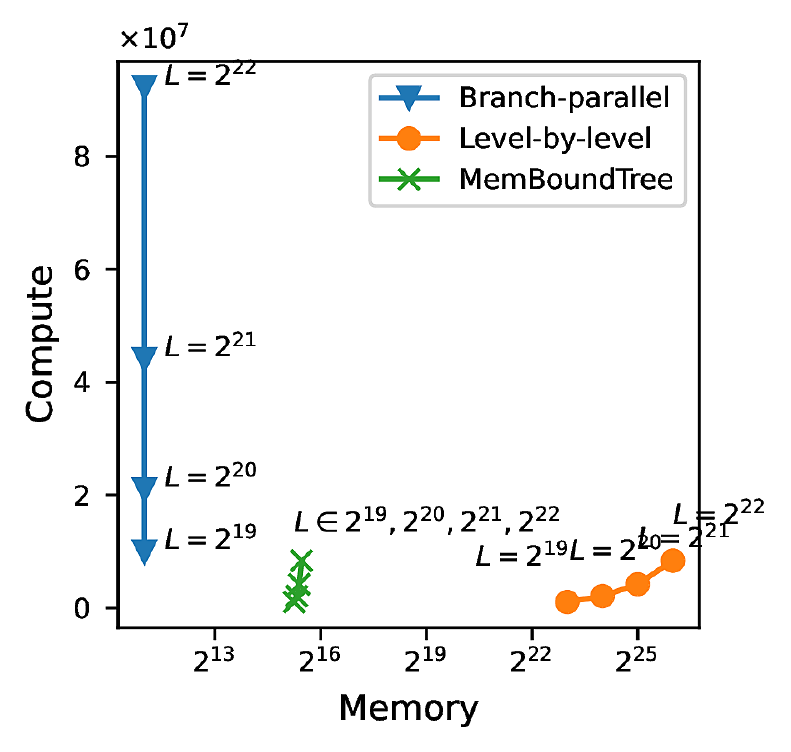}  
\vspace{-5pt}
    \caption{
    The number of PRFs evaluated (compute) and the peak memory usage (memory) for different parallelization strategies, across different table sizes (L). For both axes, lower is better. The branch-parallel approach redundantly calculates extra PRFs, while the level-by-level approach suffers from high memory usage. Our proposed approach, memory-bounded tree traversal (MemBoundTree), simultaneously performs less work while requiring much less memory -- \rev{MemBoundTree can significantly (i.e., up to 10x) improve performance by reducing memory consumption and allowing the use of larger batch sizes, which increases utilization}.
    %\# PRF calls vs memory usage for different parallelization strategies across various table sizes ($L$). Both branch-parallel and level-by-level approaches to DPF parallelization suffer from a stark tradeoff between PRF calls and memory usage. Specifically, the branch-parallel approach, while efficient in terms of memory usage, requires far more redundant computations of PRFs, limiting computational efficiency. The level-by-level approach, on the other hand, only computes each PRF once, but requires considerable memory capacity, reducing scalability by limiting the table sizes ($L$) that can be parallelized. The memory bounded tree traversal approach (MemBoundTree) achieves the best of both worlds by requiring minimal PRF operations while requiring less memory usage, enabling both scalable and efficient DPF parallel evaluation.
    }
    \label{fig:redundant}
%\end{subfigure}
\end{figure}

\begin{figure}[t]
\centering
\begin{subfigure}{.24\textwidth}
    \centering
    \includegraphics[width=1\linewidth]{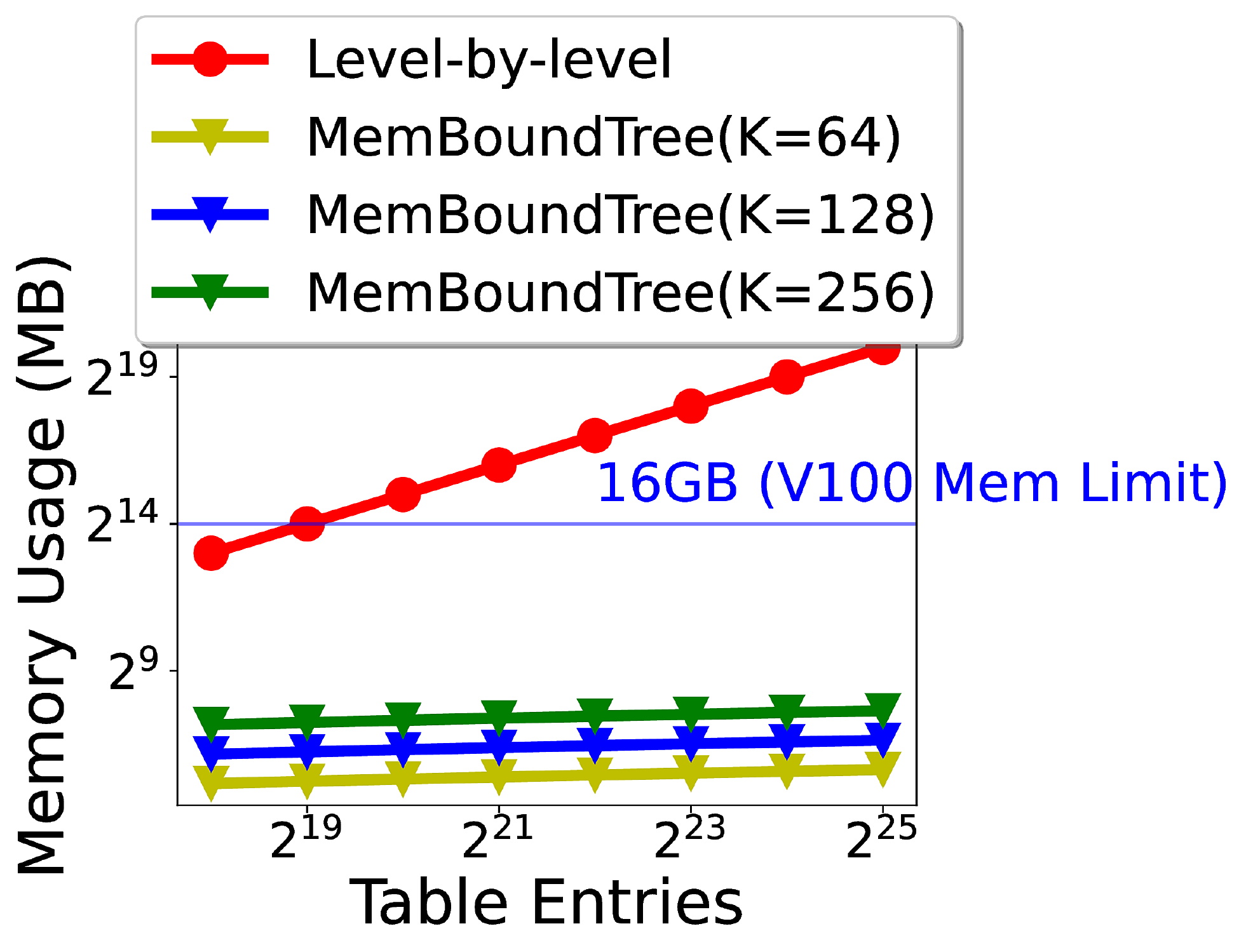}  
    \caption{Memory Usage}
    \label{fig:mem_table}
\end{subfigure}
\begin{subfigure}{.22\textwidth}
    \centering
    \includegraphics[width=.99\linewidth]{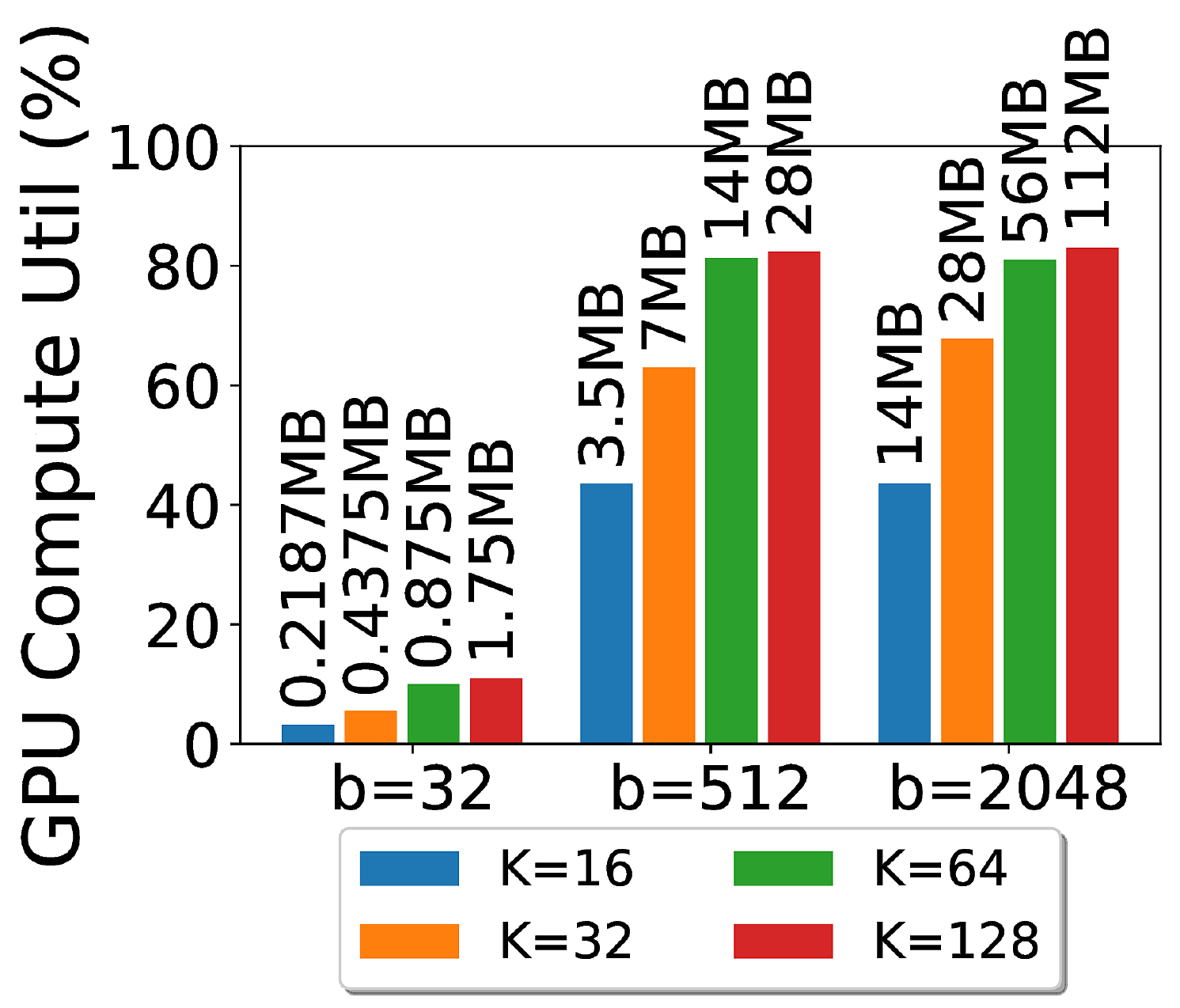}  
    \caption{GPU Utilization vs K}
    \label{fig:util_k}
\end{subfigure}
\vspace{-5pt}
\caption{The memory usage and the compute resource utilization of the memory-bounded tree traversal.}
%(a), and high resource utilization with K=128 within reasonable memory limits (b).}
\end{figure}

Memory-bounded tree traversal performs a depth-first evaluation of the DPF tree, with chunks of $K$ nodes evaluated at once in parallel for each level (Figure \ref{fig:mem_bound_tree_traversal_plus_fusion}a).
Unlike the level-by-level approach that computes and saves \emph{all} nodes in each level, the new approach only evaluates $K$ nodes per level, then immediately re-uses these node outputs by recursively computing the nodes at the next level that require these outputs, and subsequently discarding the previous node outputs. Thus, at each level, only $K$ more nodes need to be cached to memory. Hence, this approach reduces memory overhead from $O(BL)$ to $O(BKlog(L))$, making the memory overhead affordable even for large tables ((Figure \ref{fig:mem_table})). $K$, which is a hyperparameter that determines how many nodes to expand in parallel, must be large enough to expose sufficient parallelism but small enough to avoid out-of-memory complications. We empirically set $K=128$, which balances compute utilization and memory usage on a V100 GPU (Figure \ref{fig:util_k}). 
Memory-bounded tree traversal achieves both optimal work and low memory usage (Figure~\ref{fig:redundant}). \rev{As a result of achieving optimal work, low memory usage, and maximizing parallelism, the memory-bounded tree traversal method can scale to larger batch sizes and hence increase throughput and utilization up to an order of magnitude greater than a naive level-by-level approach. The memory advantage of the memory-bounded tree traversal approach is  depicted in Figure \ref{fig:redundant}, and achieves utilization benefits of a considerably larger batch size as depicted in Figure \ref{fig:util_batch}.}

\subsubsection{DPF and Matrix-Multiplication Operator Fusion} % +E +M

After evaluating the DPF, the server needs to perform a matrix multiplication between the large table and the DPF output (Section~\ref{sec:dpf}).
If we naively compute the entire output before performing a matrix multiplication, the memory must hold the entire output of the DPF and requires $O(BL)$ space. To keep the memory overhead to $O(BKlog(L))$, we \emph{fuse} the DPF evaluation operator with the matrix multiplication operator (Figure \ref{fig:mem_bound_tree_traversal_plus_fusion}b). Upon reaching a leaf node, a thread immediately performs a dot product between the table entry and the corresponding leaf node output of size $K$, accumulating the result in local memory. At the end, threads in a single thread-block coordinate to perform a cross-thread sum of the local registers to obtain the final result, using tree-summation. Fusing DPF has additional performance benefits as it reduces the number of accesses to global memory and allows interleaving between matrix-multiplication and DPF computation.

\subsubsection{Batch and Table-Size Aware Scheduling} % +E +M +C
\begin{figure}[t]
\centering
\begin{subfigure}{.22\textwidth}
    \centering
    \includegraphics[width=.99\linewidth]{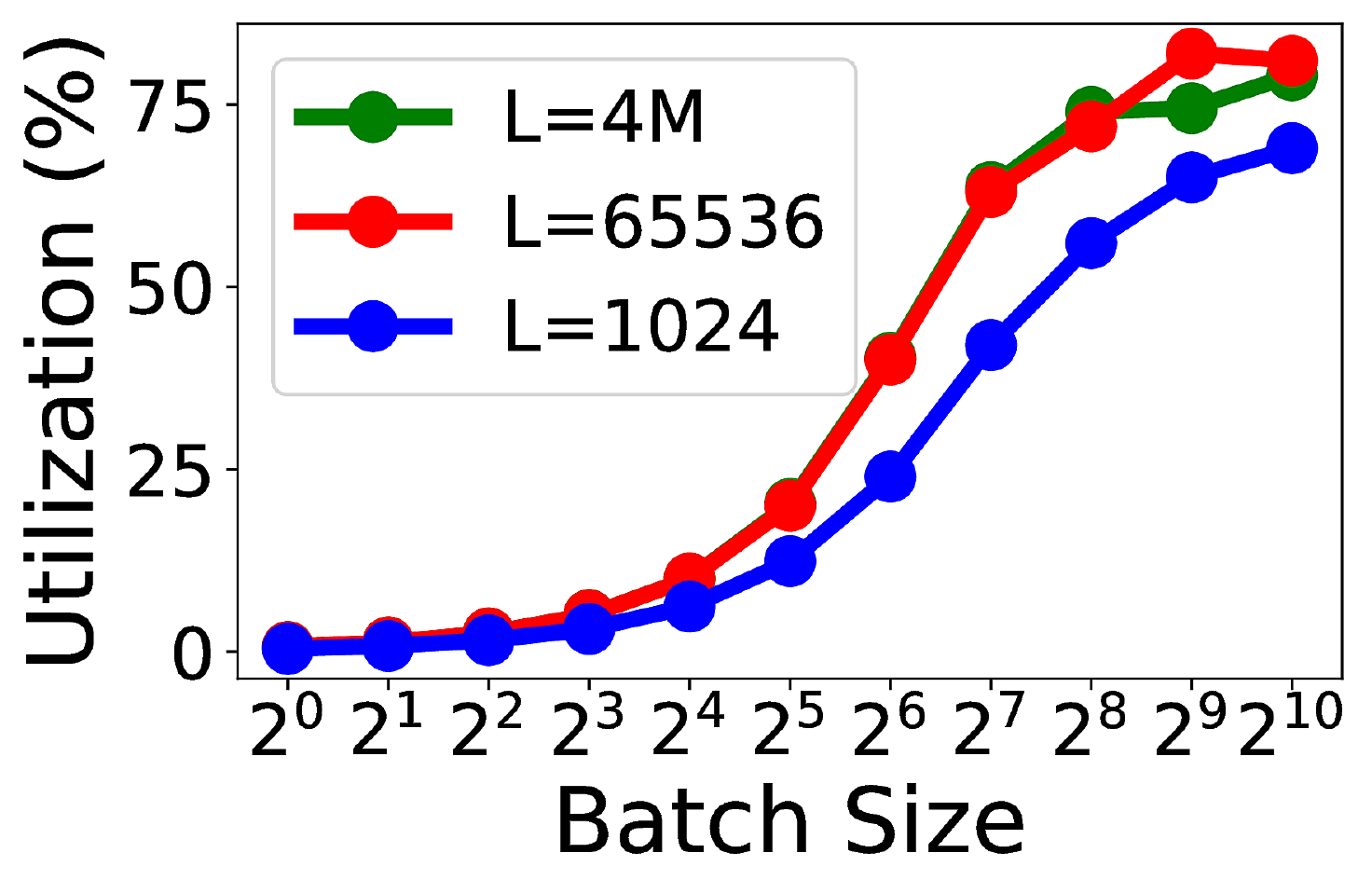}  
    \caption{Batch size vs Util}
    \label{fig:util_batch}
\end{subfigure}
\begin{subfigure}{.22\textwidth}
    \centering
    \includegraphics[width=.99\linewidth]{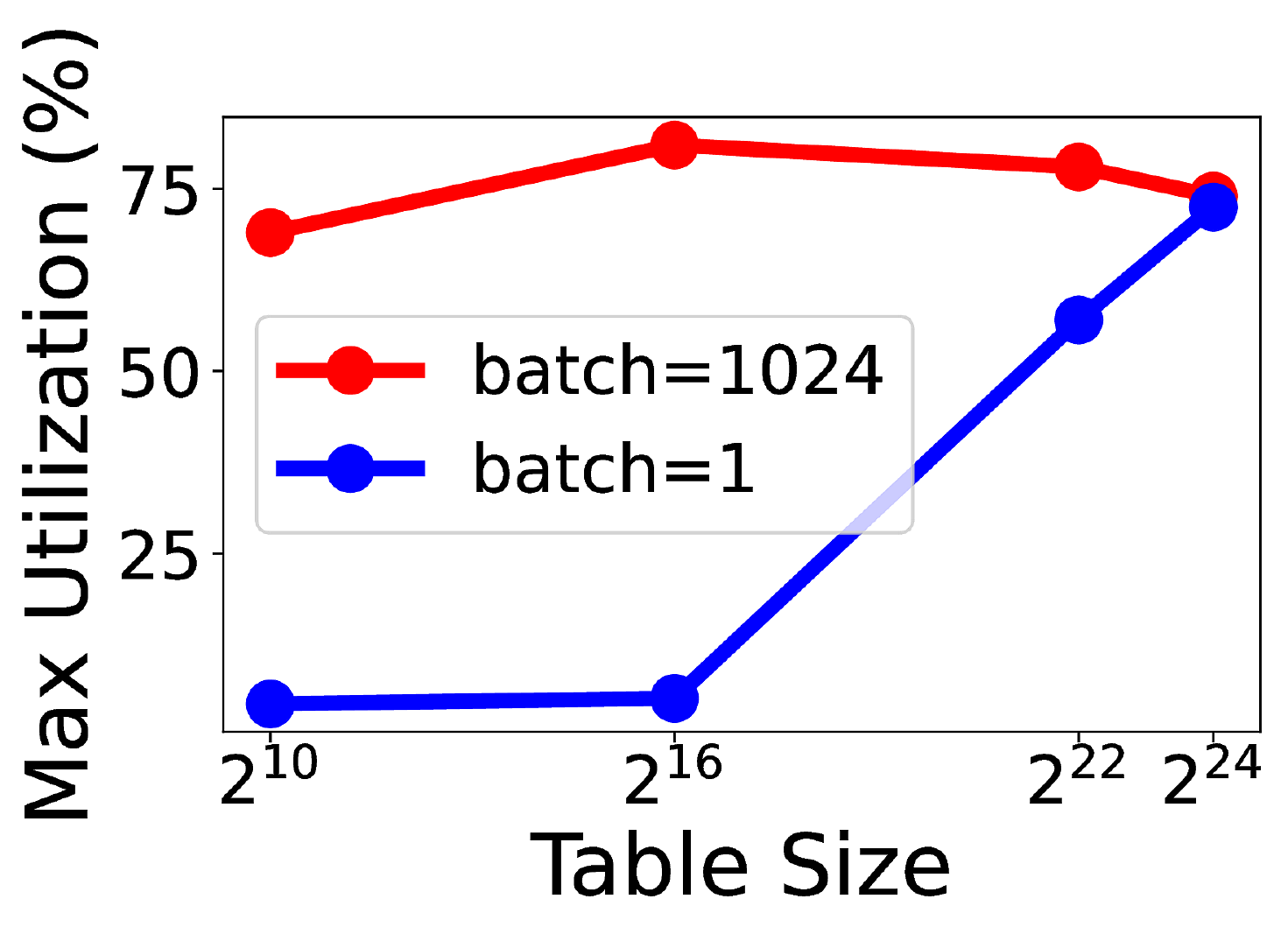}  
    \caption{Table size vs Util}
    \label{fig:util_table}
\end{subfigure}
\vspace{-5pt}
\caption{Effect of batch size (a) and table size (b) on GPU utilization. For figure (b), batch=1 utilizes cooperative groups to coordinate all available GPU resources towards computing a single DPF.}
\end{figure}

On large tables ($>2^{22}$ entries), we observe that a single DPF (batch size of 1) may have enough parallelism to sufficiently saturate GPU resources.
%, as towards the leaf nodes of the DPF-tree the number of nodes that may be computed in parallel tends towards the table size $L$ (in contrast with earlier levels towards the root for which there is less parallel work).
Hence, for very-large tables, it is preferable to use all GPU resources for the computation of a single DPF at a time, which significantly reduces latency, rather than perform batched-execution.
%which decreases latency while only attaining the same throughput.
We additionally develop a parallelization strategy based on cooperative groups \cite{coop_groups} to coordinate all GPU blocks when computing a single DPF. \rev{This single-batch strategy is selectively applied} only when the table size is very large. Figure~\ref{fig:util_table} shows that using cooperative groups with a batch size of 1 can indeed achieve high GPU utilization on extremely-large tables (with a \rev{lower} latency, which is not shown), while  it suffers from low resource utilization if incorrectly applied to smaller tables. We empirically use a threshold of $2^{22}$ entries to choose between batched execution and cooperative groups.
%Hence our selection of which parallelization approach to use depends on table size, and depending on the table size we dynamically select the appropriate strategy to deploy (batched-execution or cooperative-groups).

\subsubsection{GPU-Aware PRF Selection}
CPUs typically come with built-in hardware for popular PRFs such as AES and SHA-256 (e.g., AES-NI instructions). AES is a natural choice for the PRF on a CPU given built-in CPU hardware primitives.
However, unlike CPUs, GPUs do not offer hardware acceleration for cryptographic primitives. As a result, AES computation on a GPU is far more computationally expensive compared to a CPU. Hence, a more careful PRF selection has the potential to provide higher performance on a GPU.
In this context, we evaluate multiple PRFs including block ciphers (AES), hash functions (SHA-256), stream ciphers (ChaCha20), and others. 
We mainly show results of PIR performance based on AES-128 to match the standard PRFs used in the CPU PIR baseline.
However, we found that PRF selection has a significant impact on GPU PIR performance, and we report these results in the evaluation as well. 
Particularly, Chacha20, which is a standard stream cipher used in TLS \cite{tls_chacha}, provides noticeable performance gains.
Other non-standard PRFs, such as SipHash, can provide even more speed-up, but \rev{their security assurance may be weaker as they are not yet widely analyzed or proven in practice.} One must consider the performance and security tradeoff of a PRF to determine whether that PRF is suitable for the application at hand.

\rev{
\subsubsection{Note on Scaling to Multiple GPUs}
We note that our DPF execution strategies may be applied to multiple GPUs in the case where a single embedding table is too large to fit in a single GPU's memory. A single DPF can be computed across multiple GPUs by having each of the $N$ GPUs evaluate the DPF on a subset of the table indices, then summing the result across GPUs at the end. This approach works because  the final DPF reduction operation (a summation) is linear. Hence, we can linearly scale our DPF execution strategies across multiple GPUs by simply dividing the work in an embarrassingly parallel approach. We note that, in this scenario, each GPU effectively evaluates a DPF on a table of size $\frac{L}{N}$, hence, performance is the same as if evaluating a DPF on a smaller table size. 
Additionally, with more GPUs, a larger batch size may be needed to fully utilize GPU compute resources since the table sizes are proportionally smaller. 
Thus, for multi-GPU execution, it becomes more important to maximize batch size by using the memory-bounded tree traversal execution strategy, and a cooperative-groups approach would be less effective. 
}

\section{Accelerating Batch-PIR with ML Co-Design}

Many recommendation/language models require multiple lookups to the same embedding table. For example, recommendation models may lookup the same table tens of times to perform a single inference~\cite{deeprecsys} (e.g., a user can have multiple clicked items, if the clicked-item history is used as a feature). Multiple lookups linearly increase the cost of PIR as simple DPF-PIR only retrieves one entry at a time.

%\textcolor{red}{Kiwan: This is an interesting research question that should have been mentioned before.}

\rev{To support multiple tabe lookups more efficiently, we} adopt partial batch retrieval (PBR) \cite{multihot}, an algorithm that accelerates the retrieval of multiple entries. PBR comes at a cost; with some probability (when multiple queries map to the same internal bin), queries are dropped, which may negatively affect model quality. Hence, we co-design PBR with ML inference to improve system performance while maintaining the model quality.

%Co-designing PIR with the underlying ML application is essential towards balancing system performance with ML model quality. Below, we provide a brief background of batch-PIR techniques, then proceed to describe how to co-design these techniques with the ML application.

\subsection{Background: Batch Private Information Retrieval}

\begin{figure}[!tb]
\centering
\includegraphics[width=.9\linewidth]{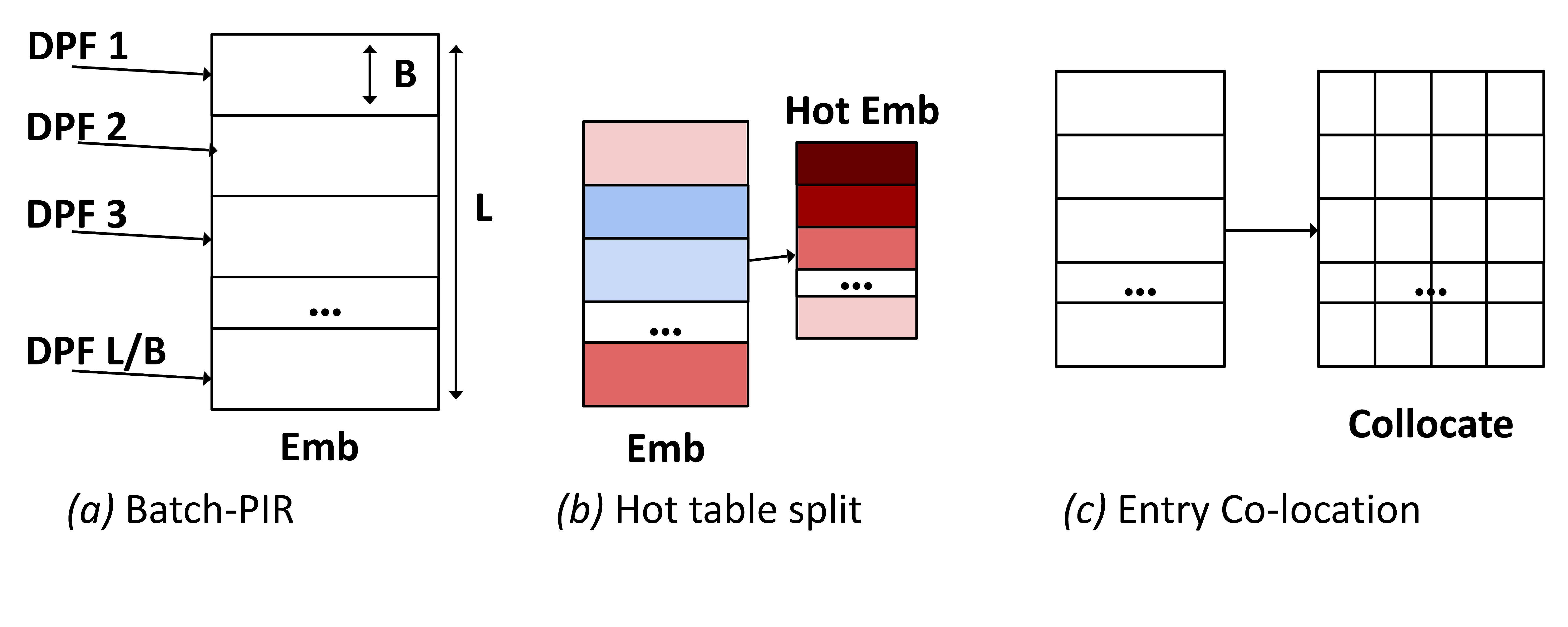}  
\iffalse
\begin{subfigure}{.15\textwidth}
    \centering
    \includegraphics[width=.75\linewidth]{figures/pir-batch.pdf}  
    \caption{Batch-PIR}
    \label{fig:batchpir}
\end{subfigure}
\begin{subfigure}{.15\textwidth}
    \centering
    \includegraphics[width=.75\linewidth]{figures/pir-hot-cold.pdf}  
    \caption{Hot Splitting}
    \label{fig:hotcold}
\end{subfigure}
\begin{subfigure}{.15\textwidth}
    \centering
    \includegraphics[width=.75\linewidth]{figures/pir-collocate.pdf}  
    \caption{Entry Co-location}
    \label{fig:collocate}
\end{subfigure}
\fi
\vspace{-5pt}
\caption{Techniques used to co-design PIR + ML. a) Partial Batch Retrieval, b) splitting the table  into a smaller hot table, and c) co-locating frequently accessed entries.}
\label{fig:batch_pir_mega}
\vspace{-5pt}
\end{figure}

Batch private information retrieval (batch-PIR) is a set of techniques to retrieve multiple private entries from a single table. In this work, we adopt the method proposed in \cite{multihot}, partial batch retrieval (PBR), which operates by segmenting table $T$ into  $\frac{L}{I}$ bins of size $I$, and issuing individual DPF-PIR queries to each bin (Figure \ref{fig:batch_pir_mega}a). This approach saves computation by a factor of $\frac{L}{I}$ in the best-case scenario where the client retrieves $\frac{L}{I}$ entries that are spread across different bins. However, a single PBR can fetch only one query from each bin. If more than one query index fall into the same bin, the rest of the queries except for the one must be dropped.

This limitation leads to a complex tradeoff between the communication efficiency and the accuracy of the retrieval. A large $I$ can reduce the accuracy of the retrieval if multiple desired entries map to the same bin.  Conversely, a smaller $I$ yields fewer conflicts, but increases communication costs. This tradeoff naturally affects model quality as dropped queries affect the model's inference.

%This approach saves computation by a factor of $\frac{L}{B}$, as each DPF that is issued is expanded over a smaller region of the table. For example, assuming that there are $\frac{L}{B}$ queries, where each query falls into different bins, all private queries can be retrieved by this scheme with $O(L)$ computation overhead, whereas naively issuing a DPF to the entire table for each query results in $O(L \times \frac{L}{B})$ computation overhead. By decreasing bin size $B$, a greater number of queries can be served at the same computation overhead, provided that they fall into different bins, thereby reducing computational workload. 
%However, decreasing bin size $B$ also increases the number of DPFs issued for a batch of queries, increasing communication overhead needlessly if a bin does not contain an index for a query. Hence, in this scheme there is a tradeoff between computation and communication. Additionally, in the case that multiple queries fall into the same bin, these queries are necessarily dropped, as a single DPF can only retrieve one item per bin. The probability of a drop is directly proportional to bin size: larger bins result in more conflicts and hence more failed queries. Therefore, there is an additional tradeoff between query success rate, and computation/communication workload. In context of ML applications, there is a complex tradeoff between model quality, which is dependent on query success rate, and system performance. Optimizing these tradeoffs is the target of the methods we present below.

\subsection{Co-Designing the ML Model and Batch-PIR} 
\label{sec:codesign}
To improve batch-PIR efficiency while minimizing effect of retrieval failures, we propose PIR-ML co-optimizations that improve the tradeoff between model accuracy and  performance.

\noindent \textbf{Frequency-Based Hot Table Split}
%\subsubsection{Frequency-Based Hot/Cold Splitting}
Many ML applications access embedding tables following a power-law distribution, where a small number of \emph{hot} indices account for the majority of  lookups~\cite{architectural_implication_recsys, zipf}. We leverage this observation and add a small \emph{hot table} that holds the top-$K$ frequently accessed indices in addition to the large \emph{full table} that holds all the embedding entries (Figure \ref{fig:batch_pir_mega}b). The hot table is constructed statically using the observed statistics from the training dataset as part of a preprocessing phase ahead of model deployment, and a small hash table is placed on a client device to provide the hot table index for the categorical feature values that are in the hot table; as this hot table is designed to be small, this index mapping can reasonably reside on client devices. At inference time, a client looks up whether the index they wish to query is in the hot table, and issues two sets of keys: one set that queries the hot table and the other for the full table. 

Simply using the hot table as a traditional cache is insecure as it leaks the number of queries to the hot/full tables. To avoid this information leakage, we predetermine a fixed number of queries $Q_{hot}$ and $Q_{full}$ 
to issue to the hot and full tables, respectively, during preprocessing.
\rev{These parameters are chosen based on the historical query request patterns for the training data, balancing the impact of dropped requests / model accuracy and performance costs.} 
\rev{
The queries issued to the hot table benefit from the lower PIR cost for accessing the small table rather than a large full table.
}
\rev{We emphasize that this design is necessary to eliminate data leakage through the number of queries that a user issues to each table.
For example, the number of queries to the hot table can reveal whether the user accesses the indices that are in the hot table.
The total number of table entries that a user accesses in both hot and full tables may also leak private information such as the number of items purchased, the number of websites visited, etc.
To remove such information leakage through the number of accesses to each table, for each inference, we require a user to issue exactly $Q_{hot}$ and $Q_{full}$ queries to the tables. 
If the user needs to read more table entries than the allocated budget, these requests are dropped; the dropped requests may impact model accuracy. 
If the user has fewer queries, then dummy queries are added to ensure that the user makes the fixed number of PIR requests.}

\noindent\textbf{Access Pattern-Aware Embedding Co-location}
Embedding table access patterns in ML applications tend to exhibit co-occurrence \cite{merci, lm_coocur} as some indices are often accessed together in a single ML inference. We co-locate \rev{the entries that are} frequently accessed \rev{together} in the same row of the table so that a single query can retrieve multiple embeddings that might be accessed together (Figure~\ref{fig:batch_pir_mega}c).
Co-location is done by profiling the training dataset and co-locating the top-$C$ embeddings that are most frequently retrieved with each embedding. 
$C$ is empirically selected. In the best-case scenario, co-location can reduce the number of queries by $C+1$.

\rev{
\textbf{Co-design Parameter Selection}
The parameters involving these two co-design techniques (frequency-based hot table splitting and embedding co-location), which involve parameters such as $Q_{hot}$, $Q_{full}$, $C$, and bin-size, as well as kernel parameters such as DPF execution batch size and DPF execution strategy are selected after sweeping the parameter space using grid search and evaluating the corresponding performance (i.e: communication and computational costs, as well as accuracy) for the target application. Note we separate training and test datasets, selecting parameters based on the training dataset, and showing results on the test dataset. Broadly, our experimental results show the pareto frontier of the performance achieved across a complete sweep of the parameter space. Generally, across applications, we found that a good choice of $Q_{hot}$ is typically 10\%-20\% of the size of the full embedding table. On the other hand, a good choice of $C$, the number of entries to collocate, depended on the application: a higher $C$ at around 4-5 (i.e: more collocation) was more beneficial for the language model task, as words in a sentence have natural associations, whereas a lower $C$ at 1-3 was better for the recommendation application. A good choice of bin size and other parameters such as DPF execution batch size and strategy, generally vary and depend on performance or accuracy constraints which may be imposed by service expectations. In summary, our co-design and kernel parameters are determined by performing a grid search across the space of possible parameters in order to find parameters that balance computation, communication and model accuracy. 
}

\rev{
\noindent\textbf{Changes to Embedding Table} Updates to the embedding table (i.e., updates/insertions/deletions) may occur over time as embedding tables can change when the model is re-trained. 
Note that updates to table entries without changing indexing (no insertion/deletion) can be done under the hood (transparent to the clients) by updating the table entries on both servers. From the client perspective, the tables are read-only. 
Full updates of embedding tables that include deletions and insertions, on the other hand, require the indexing functions on the client to be also updated. An updated hash table for the hot table needs to be sent to the client. If the full table size is changed, the hash function for indexing the full embedding table is also updated on the client.
However, this cost of a full update is only incurred when the model itself is changed or fully re-trained, which is infrequent for typical recommendation models or language models.
%Another method is to simply never (or seldom) delete an embedding entry and only append a new entry to the tables, sending the newly updated entry indices to the client device. This may considerably reduce the cost of re-indexing. 
In this paper, we study the overhead of our system assuming that full embedding table updates are infrequent enough. 
More efficient handling of table updates for other use cases that require frequent updates is left as future work.}

\section{Evaluation}

\begin{figure*}[!htb]
%\centering
%\begin{subfigure}{.23\textwidth}
%    \centering
%    \includegraphics[width=.95\linewidth]{figures/lm_final_abl.pdf}  
%    \caption{Next Word Prediction}
%    \label{}
%\end{subfigure}
%\begin{subfigure}{.23\textwidth}
%    \centering
%    \includegraphics[width=.95\linewidth]%{figures/movielens_final_abl.pdf}  
%    \caption{Movielens Recommendation}
%    \label{}
%\end{subfigure}
%\begin{subfigure}{.23\textwidth}
%    \centering
%    \includegraphics[width=.95\linewidth]{figures/taobao_final_abl.pdf}  
%    \caption{Taobao Recommendation}
%    \label{}
%\end{subfigure}
\centering
\includegraphics[width=.95\linewidth]{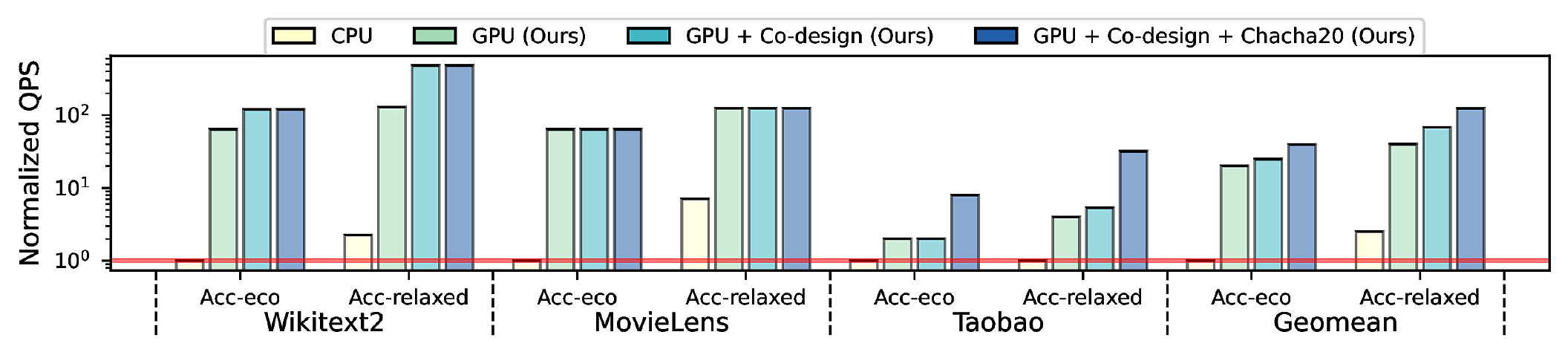} 
\vspace{-5pt}
\caption{Throughput improvement of our proposed system over the CPU baseline~\cite{google_dpr}.
While preserving accuracy (Acc-eco), our system can improve the throughput on average by 5--39$\times$. When some amount of accuracy degradation is tolerated (Acc-relaxed), the average improvement reaches 40--124$\times$.
All configurations searched within the latency ($<300$ms) and communication requirement ($<300$KB). QPS normalized by the CPU Acc-eco for each benchmark.
%
%Gains of GPU-DPF + co-design over a CPU implementation without PIR+ML co-design. Latency fixed to be $<150$ms and communication $<300$KB. Baseline model quality is shown in teal. For next word prediction, lower perplexity (ppl) is better; for recommendation models, higher area under curve (auc) is better.  1 kq/s = 1,000 queries per second.
}
\label{fig:final_ablation}
\end{figure*}

\subsection{Evaluation Setup}
\paragraph{Platforms}
We evaluate our GPU-based DPF-PIR and compare it with a state-of-the-art CPU implementation~\cite{google_dpr}. We run all GPU experiments on an NVIDIA V100 GPU, and all CPU experiments on an Intel(R) Xeon(R) Gold 6230 CPU @ 2.10GHz with 28 cores. The CPU baseline is an optimized DPF-PIR implementation from Google Research~\cite{google_dpr}, which uses AES-NI CPU hardware acceleration.

\paragraph{Datasets and Models}
We evaluate our system and the baseline by running \rev{a couple of recommendation models and a language model} on open-source datasets.
We run 
(1) a 2-layer MLP-based recommendation model~\cite{on_device_rec} with MovieLens-20M dataset~\cite{movielens}, 
(2) a 2-layer MLP-based recommendation model~\cite{on_device_rec} with Taobao Ads click/display dataset~\cite{taobao},
and (3) an LSTM model with Wikitext2 corpus~\cite{wikitext2}. 
We protect 
the user history table~\cite{din_ctr} for recommendation models 
and the word embedding for the LSTM 
using PIR.
The baseline model quality of the models we study are as follows.
For recommendation models, we use \textit{area under the receiver operating characteristic curve} (ROC-AUC or AUC) metric, where a higher AUC means better quality. Our model achieves AUC=0.7845 for MovieLens and AUC=0.58 for Taobao, similar to prior works~\cite{din_ctr, on_device_rec}.
For LSTM, we use perplexity (ppl), a measure of surprise, to measure the model quality. Following the training setup of ~\cite{wikitext2}, our model achieves ppl=92. 

% \lhl{The quality of the language model is measured using \textit{perplexity} (ppl), a measure of surprise, where lower ppl indicates better quality. We train a standard LSTM language model according to the baselines in \cite{wikitext2} replicating the baseline score of 92 \cite{wikitext2}. The recommendation models are evaluated using the \textit{area under the receiver operating characteristic curve} (ROC-AUC or AUC) metric, where a higher AUC means better quality. For recommendation models, even a 0.1\% difference in AUC are considered significant \cite{baidu_hierarchical}. For Taobao, we train an on-device model with similar architecture as in \cite{on_device_rec}, achieving similar score of 0.58 AUC; for Movielens, we likewise train a fully connected network to a similar AUC score (0.785) as in \cite{din_ctr}.}

\paragraph{System Parameters}
For application-independent experiments (Figures \ref{fig:batch_size_latency_tradeoff_salsa}--\ref{fig:cpu_compare}, Tables~\ref{tab:aes_cpu_gpu_table}--\ref{fig:prf}), unless otherwise stated, we default to an entry size of 2048 bits. Most recommendation models use entries similar or smaller than this~\cite{din_ctr, dlrm}.
Also, by default, we use a security parameter of 128 bits as standard (AES-128), and apply all proposed GPU acceleration optimizations, with a memory optimization factor $K=128$. 
%which maximizes GPU utilization as shown in the methods section).
Batch size is tuned for each experiment separately to maximize throughput while meeting latency and communication budgets (300ms and 300KB, unless stated otherwise).

%In all experiments, we conduct design space exploration across various parameters for our optimizations, with the model trained on 80\% of the public training data and evaluating model quality on 2\% of the public validation data. 

%\subsection{Throughput Improvement: Single-PIR}

%\textcolor{red}{TODO}

\subsection{End-to-End System Throughput on Applications}
\label{section_end_to_end_app}
First, we show that our proposed design \emph{significantly improves system throughput on various applications}, compared to the baseline CPU system~\cite{google_dpr}.
We evaluate key portions of our proposed design separately: 1) Applying all GPU acceleration techniques (\textbf{GPU (Ours)}), 2) Adding ML co-design (\textbf{GPU + Co-design (Ours)}), and 3) Using Chacha20 instead of AES-128 (\textbf{GPU + Co-design + Chacha20 (Ours)}).
For each design, we conducted an extensive parameter sweep across kernel hyperparameters like batch size and $K$, and across co-design hyperparameters like hot table and cold table sizes, the number of entries co-located, and the number of queries issued to each table.
We first show throughput achieved requiring a fixed model quality. Then, we additionally show throughput improvement tolerating some model quality degradation. We set the tolerated degradation to 
${<} 0.5$\% for MovieLens and Taobao and 
${<} 5$\% for Wikitext2. 

Figure~\ref{fig:final_ablation} shows that the throughput improves by \textbf{5--39$\times$} while maintaining the model quality (Acc-eco), and the improvement becomes \textbf{40--124$\times$} when small quality degradation is tolerated (Acc-relaxed). GPU optimizations account for 10--20$\times$ performance improvement, and PIR-ML co-design can additionally obtain up to 2--5$\times$ improvement. These cumulative improvements result in significant overall gains.
Co-design does not show improvement for MovieLens for this particular setup; however, the co-design is more effective for the cases with a tighter communication budget. We discuss this later in Figure~\ref{fig:tput_vs_acc_movielens}.

Table~\ref{tab:final_tab} additionally shows the unnormalized numbers for some representative points. Our proposed design improves performance from an impractical throughput (e.g., 5 QPS) to an acceptable range of hundreds of QPS.
Taobao has much higher QPS in general, because
%uses a larger table (846811 entries), 
each user queries much fewer entries per inference (2.68 on average), compared to other benchmarks (e.g., MovieLens queries 72 entries per inference on average).

\begin{table}[t]
\centering
\caption{Unnormalized QPS from Figure~\ref{fig:final_ablation}. Among our proposed design, we only show the best one (GPU + Co-design + Chacha20). Acc-eco specifies that each approach must reach the full-precision accuracy; Acc-relaxed indicates the approaches must reach within some range of full precision accuracy; see Section \ref{section_end_to_end_app}}
\label{tab:final_tab}
{\small
\begin{tabular}{c|c|c|c}
\hline
\multirow{2}{*}{Dataset} & \multirow{2}{*}{CPU} & \multicolumn{2}{c}{Ours} \\\cline{3-4}
& & Acc-eco & Acc-relaxed\\ \hline
Wikitext2 & 5 & 577 & 2,306\\ \hline
MovieLens & 44 & 2,821 & 5,476\\ \hline
Taobao & 8k & 64k & 256k\\ \hline
\end{tabular}
}
\vspace{-5pt}
\end{table} 
%

%TODO: Table 4 into another result? (e.g., throughput improvement of a single query)

\subsection{End-to-End System Latency}

\begin{figure}%[!htb]
%\centering
%\begin{subfigure}{.23\textwidth}
%    \centering
%    \includegraphics[width=.95\linewidth]{figures/lm_final_abl.pdf}  
%    \caption{Next Word Prediction}
%    \label{}
%\end{subfigure}
%\begin{subfigure}{.23\textwidth}
%    \centering
%    \includegraphics[width=.95\linewidth]%{figures/movielens_final_abl.pdf}  
%    \caption{Movielens Recommendation}
%    \label{}
%\end{subfigure}
%\begin{subfigure}{.23\textwidth}
%    \centering
%    \includegraphics[width=.95\linewidth]{figures/taobao_final_abl.pdf}  
%    \caption{Taobao Recommendation}
%    \label{}
%\end{subfigure}
\centering
\includegraphics[width=.7\linewidth]{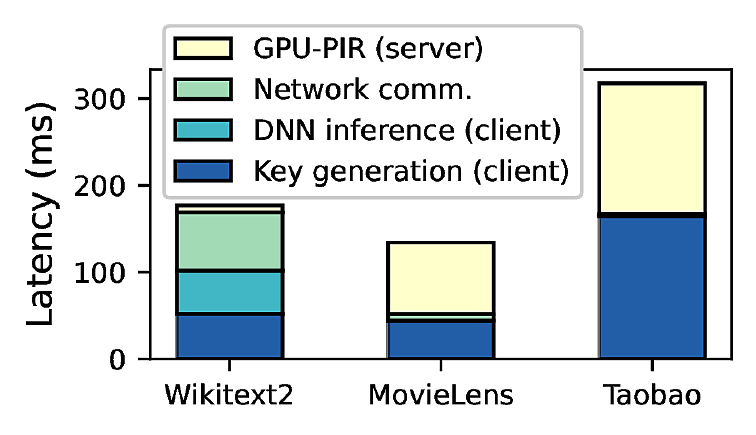} 
\vspace{-5pt}
\caption{End-to-end latency breakdown of an inference query (i.e: time from client request to receiving and computing the result). Our proposed system makes the PIR latency much lower (Wikitext2) or comparable (MovieLens, Taobao) to the latency of other components. We are able to keep \rev{end-to-end} latency within a reasonable 500 ms per inference which is acceptable in standard SLAs \cite{architectural_implication_recsys}
%Gen / Model inference is run on single-core 3. Network latency is estimated as time to transfer a batch of queries/embeddings for a single inference assuming a network bandwidth of 60 Mbit/s. While enabling privacy increases latency costs, it is kept within a reasonable range of $<500$ ms per inference.
%\label{tab:final_tab_latency
}
\label{fig:latency_breakdown}
\vspace{-10pt}
\end{figure}

We subsequently show the impact of our system on end-to-end inference latency to show that the latency overhead of our GPU-PIR results in acceptable standards for real-world applications.
Four components that affect inference latency include: (1) client-side key generation ($Gen$), (2) PIR ($Eval$; our paper's main focus), (3) client-server network communication (4) client on-device DNN inference.
We measure the latency of key generation and DNN inference directly on a single Intel Core i3 CPU. We \rev{estimate} the network latency assuming 60 Mbit/s bandwidth as in 4G networks \cite{throughput_4g}. 

Figure~\ref{fig:latency_breakdown} shows that PIR is not the sole dominating latency bottleneck anymore, costing comparable or less latency compared to other sources. While the overall end-to-end latency is much larger than a no-privacy system, the end-to-end latency still falls under the typical service level requirement (SLA) of many real-world applications \cite{architectural_implication_recsys}.

\subsection{Detailed Analysis of System Optimizations}

%\begin{figure}[!htb]
%\centering
%\begin{subfigure}{.23\textwidth}
%    \centering
%    \includegraphics[width=\linewidth]{figures/plot3_line_1.pdf}  
%    \caption{Table Entries = 1M}
%    \label{fig:gpu_opt_1m}
%\end{subfigure}
%\begin{subfigure}{.23\textwidth}
%    \centering
%    \includegraphics[width=\linewidth]{figures/plot3_line_2.pdf}  
%    \caption{Table Entries = 4M}
%    \label{fig:gpu_opt_4m}
%\end{subfigure}
%\caption{Throughput vs latency for our GPU acceleration strategy across different optimizations and table sizes, varied across batch size (table entries are fixed at 2048 bits). Our optimizations include \underline{e}liminating redundancy, \underline{m}emory optimization, \underline{c}ooperative groups. 1 kq/s = 1,000 queries per second. }
%\label{fig:batch_size_latency_tradeoff_salsa}
%\vspace{-10pt}
%\end{figure}

\begin{figure}%[!htb]
\centering
\includegraphics[width=.99\linewidth]{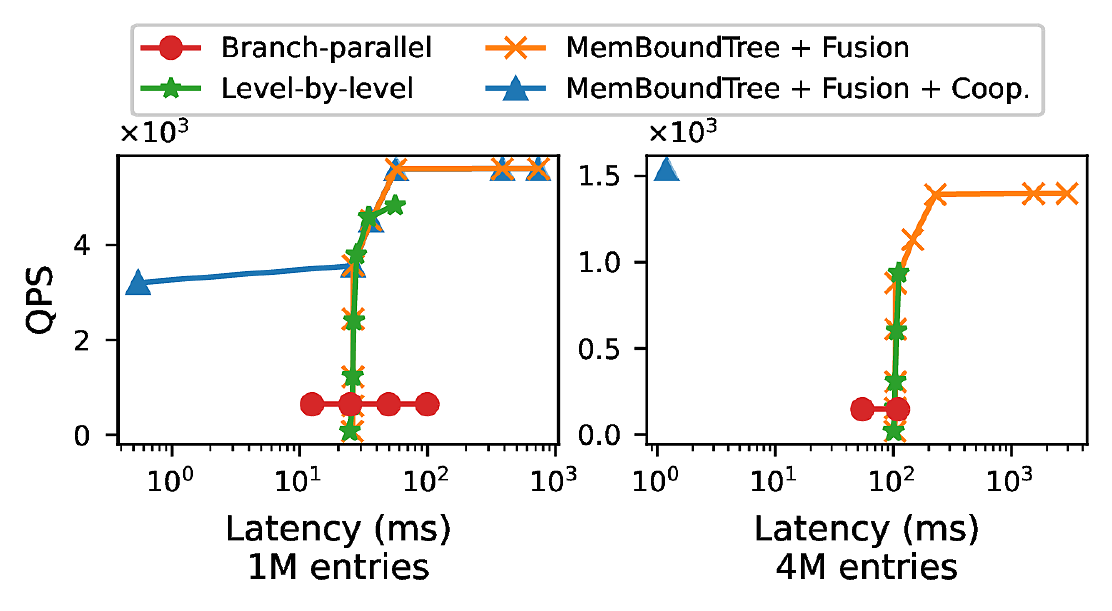} 
\vspace{-5pt}
\caption{Throughput vs latency for different GPU optimizations: branch-parallel (red), level-by-level (green), memory-bounded tree traversal and operator fusion (orange), and batch/table-size aware scheduling with cooperative groups (blue).}
\label{fig:batch_size_latency_tradeoff_salsa}
\vspace{-10pt}
\end{figure}

%\begin{table}[!htb]
%\centering
%\scalebox{.85}{
%\begin{tabular}{c|c|c|c|c}
%\hline
%Table Size & \begin{tabular}[c]{@{}c@{}}QPS\\ no fusion\end{tabular} & \begin{tabular}[c]{@{}c@{}}Latency\\ no fusion\end{tabular} & \begin{tabular}[c]{@{}c@{}}QPS\\fusion\end{tabular} & \begin{tabular}[c]{@{}c@{}}Latency\\fusion\end{tabular}\\ \hline
%16K & 193k & 1.49 & 348k (1.8$\times$) & 0.9 (1.6$\times$)\\ \hline
%64K & 49k & 5.8 & 89k (1.8$\times$) & 3.55 (1.6$\times$)\\ \hline
%1M & 3k & 92  & 5.5k (1.9$\times$) & 56.5 (1.6$\times$)\\ \hline
%4M & 759 & 369 & 1.4k (1.8$\times$) & 226 (1.6$\times$)\\ \hline
%\end{tabular}}
%\caption{Impact of fusing and interleaving DPF expansion with matrix multiplication kernels using the Chacha20 PRF. Each entry in the table is 256 bytes. Latency is ms. \textcolor{red}{Kiwan: I think this can go away. The improvement is consistent so showing it for different size is not so interesting, and Fig 14 also shows the benefit of fusion anyways.}}
%\label{fig:fusion_performance}
%\vspace{-5pt}
%\end{table}

Here, we evaluate and isolate the effects of our proposed system optimizations, starting with GPU kernel optimizations, and concluding with ML co-design optimizations.

%We evaluate our set of GPU optimizations to accelerate DPF-PIR and compare it with the state-of-the-art CPU implementation. We run all GPU experiments on a GV100 GPU, and all CPU experiments on a Intel(R) Xeon(R) Gold 6230 CPU @ 2.10GHz with 28 cores. The CPU baseline is an optimized DPF-PIR implementation from Google Research~\cite{google_dpr}, which uses AES-NI CPU hardware acceleration. Unless otherwise specified, We default to a security parameter of $\lambda = 128$ bits, with each table entry containing 256 bytes.

%\begin{figure}[!htb]
%\centering
%\includegraphics[width=.95\linewidth]{figures/plot3.pdf} 
%\caption{Effect of each optimization from Section~\ref{sec:todo} to throughput, when applied cumulatively. Redundancy-aware memoization (+Memoize) and memory-related optimizations (+Mem opt.) significantly improves the QPS, while cooperative grouping (+Coop.) only helps for very large tables. All targeting a latency budget of ${<}150$ms.}
%\label{fig:effect_optimization}
%\end{figure}

\noindent \textbf{Performance Impact of Each GPU Optimization}
%Figure~\ref{fig:effect_optimization} shows the effect of cumulatively applying optimizations from Section~\ref{sec:todo}: Redundancy-aware memoization (Section~\ref{sec:todo}; +Memoize), memory-aware tree traversal and operator fusion (Section~\ref{sec:todo}; +Mem opt.) and using a cooperative group within a single batch (Section~\ref{sec:todo}; +Coop.).
%All points were selected with tuning the batch size for the best throughput while meeting the latency budget (${<}150$ms).
%
%As shown, each successive optimization increases the throughput significantly. For small to medium tables (1M entries), applying memoization and memory-related optimizations achieves the best QPS. 
%Only for very large tables (4M entries), using a cooperative group additionally improves the throughput. 
%Adding memory optimizations additionally improves the throughput over memoization as the system can use a larger batch size. The benefit is more significant for larger tables (4M entries) as memory becomes more pressured.
%We additionally highlight some of these representative performance points in Table~\ref{tab:aes_cpu_gpu_table}.
%Note that these numbers account for both expanding the DPF and executing the subsequent matrix multiply.
Figure~\ref{fig:batch_size_latency_tradeoff_salsa} plots the latency-throughput tradeoff for each GPU optimization.
As shown, our proposed optimizations increase the latency-throughput pareto frontier significantly.
As discussed in Section~\ref{sec:bp_and_lvl}, branch-parallel (red) cannot achieve high QPS. Level-by-level (green) is much better, but still limited, as it is bottlenecked by the memory capacity. The proposed memory-bounded tree traversal and operator fusion (orange) is able to increase the throughput further when some latency degradation is tolerated, by using less memory and allowing additional batching.
For very large tables (Figure~\ref{fig:batch_size_latency_tradeoff_salsa} (right)), table-size aware scheduling with cooperative groups (blue) obtains significantly better latency without harming throughput.

\begin{figure}%[t!]
    \centering
\begin{subfigure}{.22\textwidth}
    \centering
    \includegraphics[width=.99\linewidth]{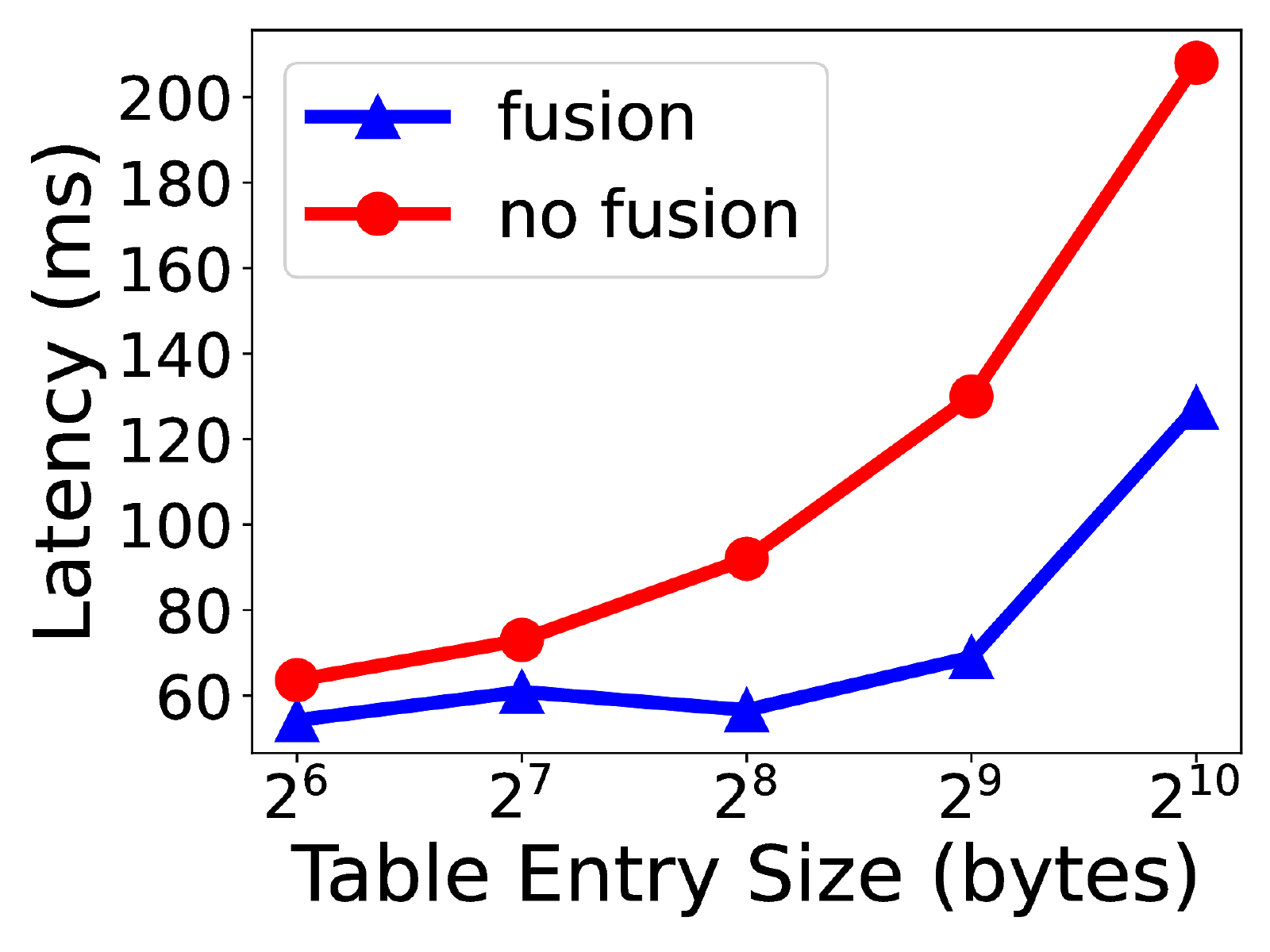}  
    \caption{Latency}
    \label{}
\end{subfigure}
\begin{subfigure}{.22\textwidth}
    \centering
    \includegraphics[width=.99\linewidth]{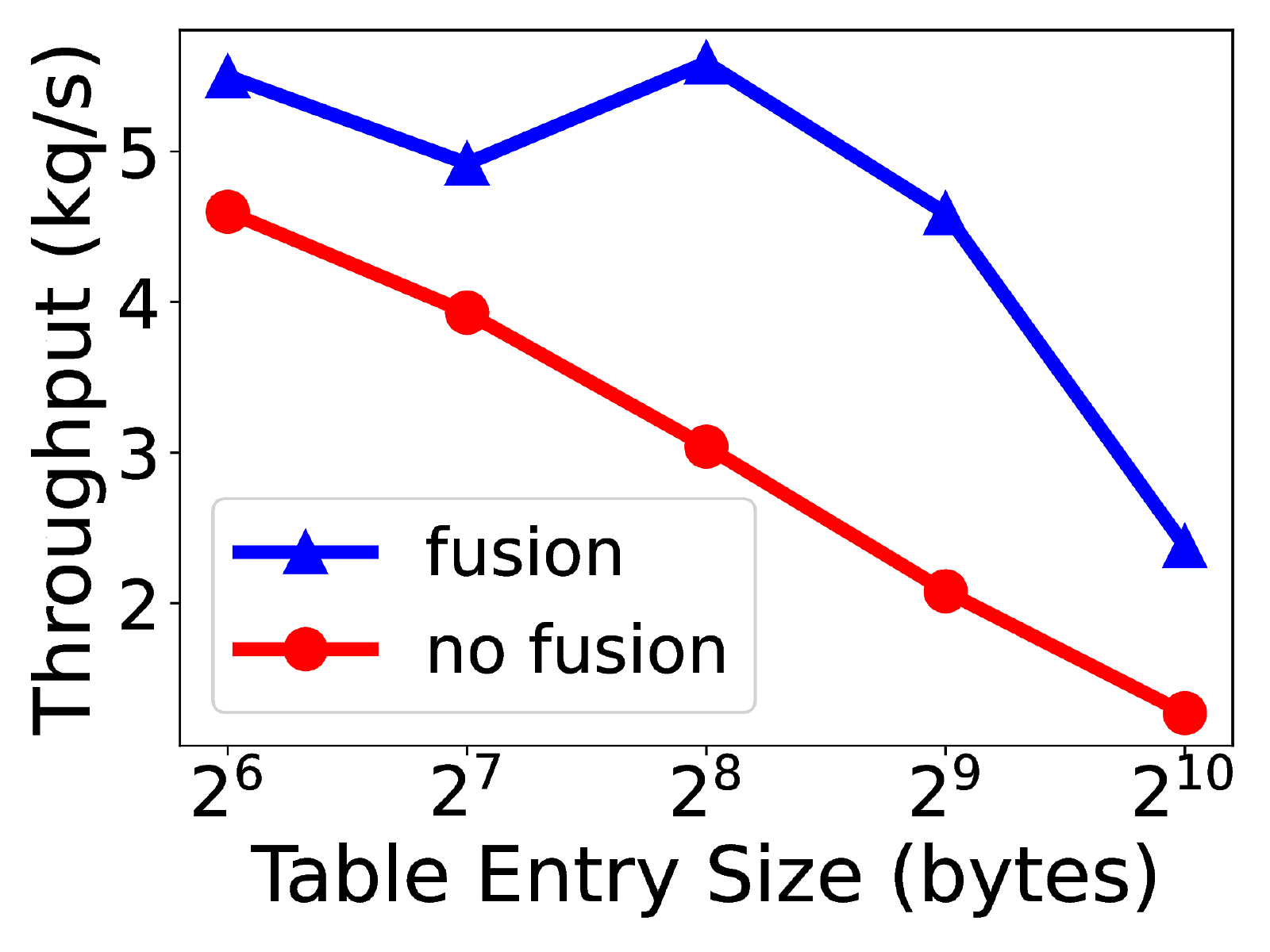}  
    \caption{Throughput}
    \label{}
\end{subfigure}
    \caption{Performance impact of table entry size on PIR performance, with and without operator fusion. }
    \label{fig:embedding_entry_size_throughput}
    \vspace{-5pt}
\end{figure}

\noindent \textbf{Performance Impact of Operator Fusion}
%\textcolor{red}{Kiwan: Do we want to still have this separately? Now fusion is part of the memory operation. Is it already contained in Figure 15?}
Figure~\ref{fig:embedding_entry_size_throughput} shows the performance benefits of fusing the subsequent matrix multiplication with DPF evaluation, across different table entry sizes. Generally, fusing and interleaving the two kernels offer significant ($> 1.5 \times$) improvements in both throughput and latency. Figure~\ref{fig:embedding_entry_size_throughput} was obtained with a table with 1M entries; however, the improvement is similar across other table sizes.
%on less compute-bound PRFs like Chacha20 as it allows more efficient scheduling of compute and memory operations.  

\noindent \textbf{Performance Impact of Embedding Entry Size}
Figure~\ref{fig:embedding_entry_size_throughput} also shows the impact of different table entry sizes on latency and throughput.
%the size of the table entries affects the amount of work done in the subsequent matrix multiplication, and hence may affect throughput and latency.
%Recall that the size of the table entries affects the amount of work done in the subsequent matrix multiplication, and hence may affect throughput and latency. We show the impact of table entry size on latency and throughput in Figure~\ref{fig:embedding_entry_size_throughput}.
Tables with entry sizes of ${<} 512$ bytes do not significantly degrade performance, especially with operator fusion. This is because the memory operations are tightly interwoven with the subsequent matrix operations with operator fusion. As the latency and throughput does not linearly degrade with increasing entry size, co-locating and retrieving multiple entries at once becomes efficient (Section~\ref{sec:codesign}).

\begin{figure}%[tb!]
    \centering
    \includegraphics[width=.75\linewidth]{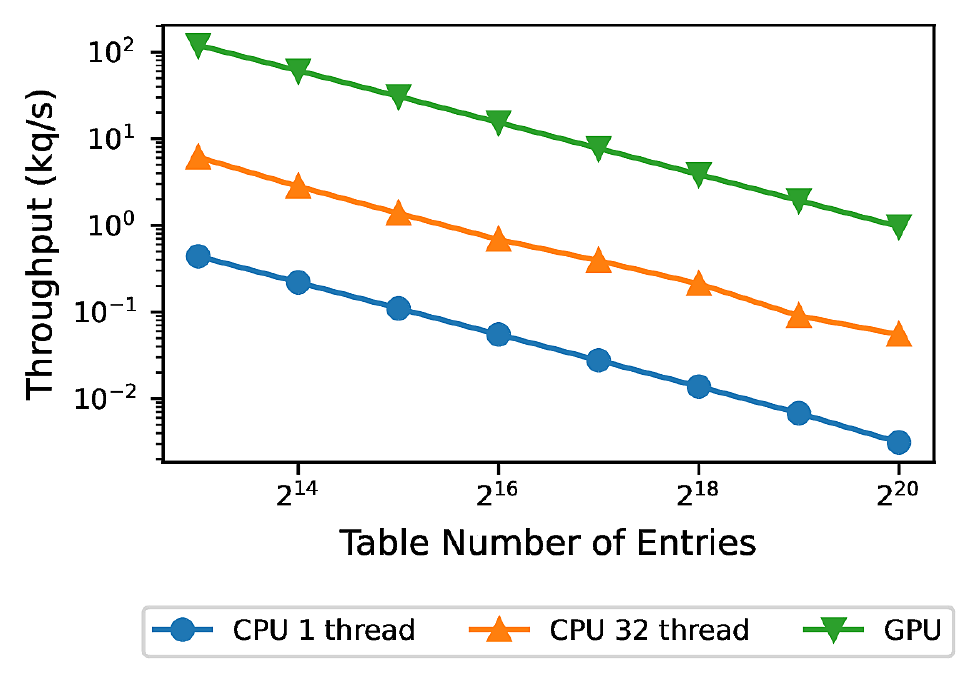}
    \vspace{-5pt}
    \caption{Comparison of throughput performance attained by GPU DPF acceleration compared to an optimized CPU baseline.  1 kq/s = 1,000 queries per second. All methods use the AES-128 PRF.}
    \label{fig:cpu_compare}
    \vspace{-5pt}
\end{figure}

\noindent \textbf{Detailed Comparison with CPU}
We compare our GPU-PIR implementation against an optimized CPU implementation from Google Research \cite{google_dpr}. Note that, Google Research's CPU implementation of DPFs uses AES-128 for its PRF, and utilizes AES-NI hardware intrinsics to accelerate PRF computation. Figure \ref{fig:cpu_compare} compares the throughput attained by the memory-efficient GPU DPF acceleration strategy against a 1-threaded and 32-threaded (fully-utilized) CPU version on different table sizes. Using  AES-128 as in the CPU DPF, our GPU implementation consistently achieves $>17 \times$ speedup over the 32-threaded CPU implementation. We show the same data in Table~\ref{tab:aes_cpu_gpu_table}.
%With a different choice of PRF, in this case Chacha20 with 12 rounds, we see over $100 \times$ speedup over multithreaded CPU execution. We show the same data in Table~\ref{tab:aes_cpu_gpu_table}. Generally, our GPU-based PIR can accelerate PIR by more than an order of magnitude over the CPU. 

% \usepackage{multirow}
\begin{table}%[htb!]
\caption{Throughput / latency comparison of our GPU acceleration (all optimizations) vs single and multi-threaded CPU implementations, on tables with an entry size of 2048 bits. Both use AES-128 as their PRF.
%, and a security parameter of 128. 
The CPU DPF baseline is taken from \cite{google_dpr} and is an optimized CPU implementation that uses AES-NI hardware intrinsics. Bytes indicates the size of the DPF key that is transferred between client and server for that table size. }
\label{tab:aes_cpu_gpu_table}
\vspace{-5pt}
\centering
{\footnotesize %scriptsize
\begin{tabular}{|c|c|c|c|c|}
\hline
\# Entries               & Bytes & Strategy      & QPS & Latency (ms) \\ \hline
\multirow{3}{*}{16K}   & \multirow{3}{*}{896}         & GPU           & 60,347                    & 3.2          \\ \cline{3-5} 
                         &                              & CPU 1-thread  & 22                       & 9            \\ \cline{3-5} 
                         &                              & CPU 32-thread & 2,810                      & .71          \\ \hline
%\multirow{3}{*}{64K}   & \multirow{3}{*}{1024}        & GPU           & 15,258                    & 18.5         \\ \cline{3-5} 
%                         &                              & CPU 1-thread  & 5                        & 36           \\ \cline{3-5} 
%                         &                              & CPU 32-thread & 688                      & 2.9          \\ \hline
\multirow{3}{*}{1M} & \multirow{3}{*}{1280}        & GPU           & 1,358                     & 1.4          \\ \cline{3-5} 
                         &                              & CPU 1-thread & 1.3                      & 638          \\ \cline{3-5} 
                         &                              & CPU 32-thread & 21.2                     & 36           \\ \hline
\multirow{3}{*}{4M} & \multirow{3}{*}{1408}        & GPU           & 468                      & 4.18         \\ \cline{3-5} 
                         &                              & CPU 1-thread  & 0.78                      & 2579.8       \\ \cline{3-5} 
                         &                              & CPU 32-thread & 12                       & 160.1        \\ \hline
\end{tabular}}
\end{table}

\begin{table}%[t!]
\centering
\caption{Performance evaluation of memory-efficient GPU DPF with different PRF functions, on a table of size 1,048,576, with batch size 512, and a security parameter of 128 bits.}
\label{fig:prf}
\vspace{-5pt}
{\footnotesize %\scriptsize
\begin{tabular}{c|c|c|c}
\hline
PRF         & \begin{tabular}[c]{@{}c@{}}Type\end{tabular}           & \begin{tabular}[c]{@{}c@{}}Latency (ms)\end{tabular} & \begin{tabular}[c]{@{}c@{}}QPS\end{tabular} \\ \hline
AES-128     & \begin{tabular}[c]{@{}c@{}}Block Cipher  (Ctr Mode)\end{tabular} & 591                                                     & 965                                                               \\ \hline
SHA-256     & \begin{tabular}[c]{@{}c@{}}Hash  (HMAC)\end{tabular}             & 659                                                     & 921                                                               \\ \hline
Chacha20  & \begin{tabular}[c]{@{}c@{}}Stream Cipher\end{tabular}           & 174                                                     & 3,640                                                              \\ \hline
%Chacha20-12  & \begin{tabular}[c]{@{}c@{}}Stream Cipher\end{tabular}           & 113                                                     & 5,598                                                              \\ \hline
%Salsa20-8   & \begin{tabular}[c]{@{}c@{}}Stream \\ Cipher\end{tabular}           & 82.7                                                    & 7499                                                              \\ \hline
SipHash     & PRF                                                                & 82.3                                                    & 7,447                                                              \\ \hline
HighwayHash & PRF                                                                & 320                                                     & 1,973                                                              \\ \hline
\end{tabular}}
%\vspace{-3pt}
\end{table}

\noindent  \textbf{Performance Impact of PRF}
Table~\ref{fig:prf} shows the performance of using different PRF functions on a table with 1M entries, a batch size of 512, and a security parameter of 128-bits. Lightweight PRFs can significantly improve the GPU-PIR performance over AES-128.
In particular, Chacha20, a well-accepted PRF that is used in high-security applications including TLS 1.3~\cite{tls_chacha}, improves the latency and throughput significantly compared to AES-128. 
Other lightweight PRFs can improve the throughput even more if their security is acceptable for the target use case.

%For example, the Chacha20 stream ciphers with different numbers of rounds provide good performance while appropriate security. While Chacha20 with 8 rounds has been broken \cite{salsa_broke}, Chacha20 with 12 and 20 rounds has no known attack. Chacha20 with 20 rounds is used in TLS 1.3, and is a well accepted PRF for applications with high security requirements \cite{tls_chacha}. 
%Thus, choice of PRF has significant performance implications and must be chosen depending on the security and performance requirements of the target application.

%\begin{figure*}[h!]
%\centering
%\begin{subfigure}{.33\textwidth}
%    \centering
%    \includegraphics[width=.95\linewidth]{figures/lm_throughpt_ppl.pdf}  
%    \caption{Next Word Prediction}
%    \label{}
%\end{subfigure}
%\begin{subfigure}{.33\textwidth}
%    \centering
%    \includegraphics[width=.95\linewidth]{figures/movielens_out_joined_throughpt_auc.pdf}  
%    \caption{Movielens Recommendation}
%    \label{}
%\end{subfigure}
%\begin{subfigure}{.33\textwidth}
%    \centering
%    \includegraphics[width=.95\linewidth]{figures/taobao_out_joined_throughpt_auc.pdf}  
%    \caption{Taobao Recommendation}
%    \label{}
%\end{subfigure}
%\caption{System throughput vs model quality with and without co-design across applications on a single V100 GPU. Communication is fixed to be ${<}300$KB per inference, and latency to be ${<}150$ms. ppl: lower is better; auc: higher is better.  1 kq/s = 1,000 queries per second.}
%\label{fig:throughput_vs_acc}
%\vspace{-10pt}
%\end{figure*}

\subsection{PIR + ML Co-Design}

\begin{figure}[t]
\centering
\begin{subfigure}{.23\textwidth}
\includegraphics[width=\linewidth]{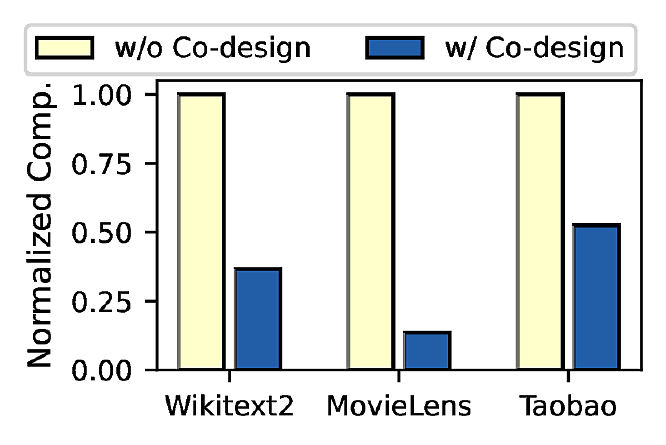} 
\caption{Computation overhead
}
\label{fig:comp_vs_model}
\end{subfigure}
\begin{subfigure}{.23\textwidth}
\includegraphics[width=\linewidth]{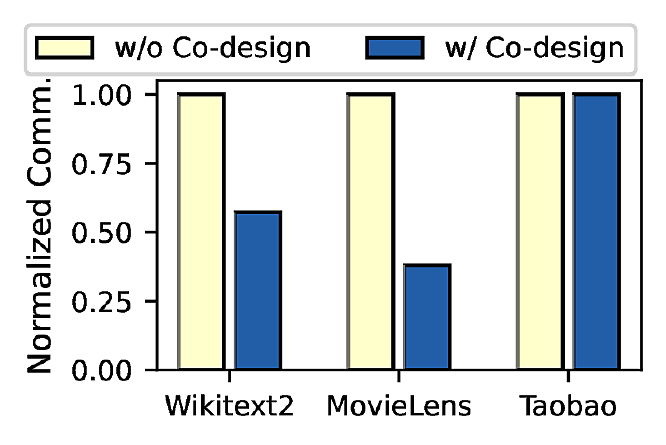} 
\caption{Communication overhead
}
\label{fig:comm_vs_model}
\end{subfigure}
\vspace{-5pt}
\caption{Computation (a) and communication (b) needed to achieve a target model accuracy (Acc-relaxed from Figure~\ref{fig:final_ablation}), with and without ML co-design. Co-design improves computation overhead by 1.9--7.4$\times$ and communication overhead by 1--2.6$\times$.
}
\label{fig:tradeoffs}
\vspace{-5pt}
\end{figure}

Private on-device ML inference often requires the private retrieval of a batch of embeddings from the same table. We evaluate our techniques that co-design ML inference and batch PIR, and demonstrate how our co-design techniques significantly improve model quality vs system performance tradeoffs.

%\begin{figure*}[htb!]
%\centering
%\begin{subfigure}{.23\textwidth}
%    \centering
%    \includegraphics[width=.95\linewidth]{figures/lm_communication_vs_computation.pdf}  
%    \caption{Next Word Prediction}
%    \label{}
%\end{subfigure}
%\begin{subfigure}{.23\textwidth}
%    \centering
%    \includegraphics[width=.95\linewidth]{figures/movielens_communication_vs_computation.pdf}  
%    \caption{Movielens Recommendation}
%    \label{}
%\end{subfigure}
%\begin{subfigure}{.23\textwidth}
%    \centering
%    \includegraphics[width=.95\linewidth]{figures/taobao_communication_vs_computation.pdf}  
%    \caption{Taobao Recommendation}
%    \label{}
%\end{subfigure}
%\caption{Pareto curve of tradeoff between communication with model accuracy fixed to be within 2\% of the baseline.}
%\label{fig:comm_vs_comp}
%\vspace{-10pt}
%\end{figure*}

%\noindent \textbf{Computation vs Model Quality} 
\noindent \textbf{Computation Savings}
%We show the tradeoff between computation and model quality given a fixed communication limit. 
%Figure \ref{fig:comp_vs_model} shows this tradeoff across various applications when communication is fixed to be less than 300KB. As shown, the co-design with both a hot table and embedding entry co-location obtains up to 2--3$\times$ improvement in computation at a fixed model quality.
Figure~\ref{fig:comp_vs_model} shows the computation needed to reach a target accuracy with and without ML co-design.
We fixed the communication below $300$KB, and target Acc-relaxed from Figure~\ref{fig:final_ablation}.
Figure~\ref{fig:comp_vs_model} shows that co-design reduces the computation significantly, by 1.9$\times$--7.4$\times$.
%Taobao's communication overhead was already too small (${<} 3$KB) and did not improve.
%Co-design can be especially useful when the communication is expensive, e.g., when using 3G/4G network.

%\noindent \textbf{Communication vs Model Quality}
\noindent \textbf{Communication Savings}
%We show the tradeoff between communication and model quality given a fixed computation limit. 
Figure~\ref{fig:comm_vs_model} shows the communication needed to reach a target accuracy (Acc-relaxed) with and without ML co-design. We fixed the computation to be less than 100K PRFs per batched inference for Wikitext2 and MovieLens, and 5M PRFs for Taobao.
With a fixed computation budget, the result shows that co-design improves the communication overhead by 1.7$\times$ and 2.6$\times$ for Wikitext2 and MovieLens, respectively. Taobao's communication overhead was already too small (${<} 3$KB) and did not improve.
Co-design can be especially useful when the communication is expensive, e.g., when using 3G/4G network.

\begin{figure}[t]
\centering
\includegraphics[width=.99\linewidth]{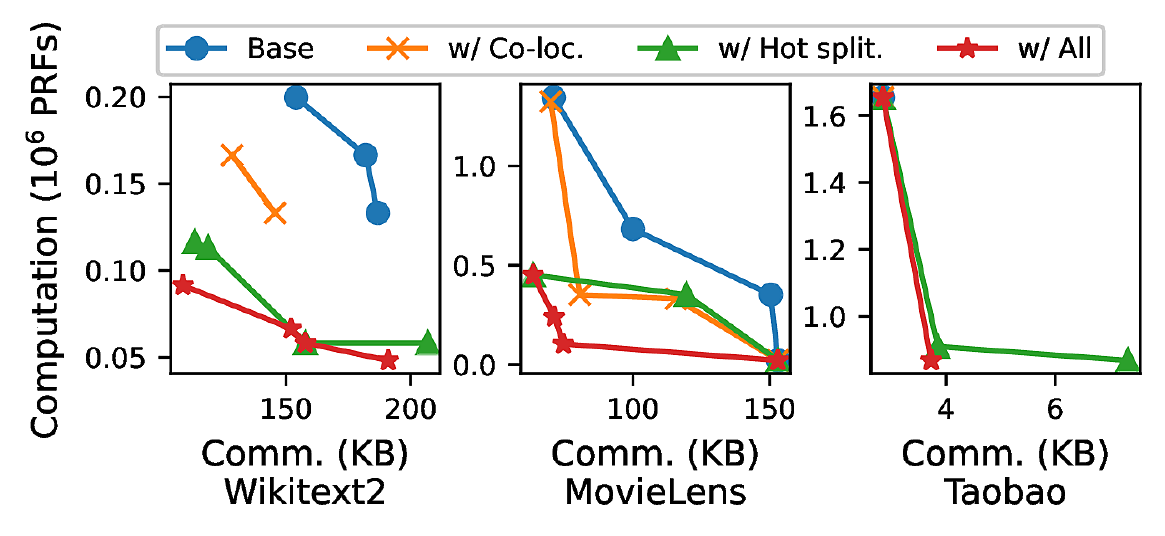} 
\vspace{-5pt}
\caption{Pareto curve of tradeoff between computation and communication with model accuracy fixed to be within 2\% of the baseline.}
\label{fig:comm_vs_comp}
%\vspace{-10pt}
\end{figure}

\noindent \textbf{Communication vs Computation}
We show the tradeoff between computation and communication with the fixed model quality. Figure \ref{fig:comm_vs_comp} shows this tradeoff across various applications, with model quality fixed to be within $2\%$ of the full precision baseline. Co-design optimizations obtain significantly better tradeoffs than plain batch-PIR. 

\begin{figure}%[t]
\centering
\includegraphics[width=.8\linewidth]{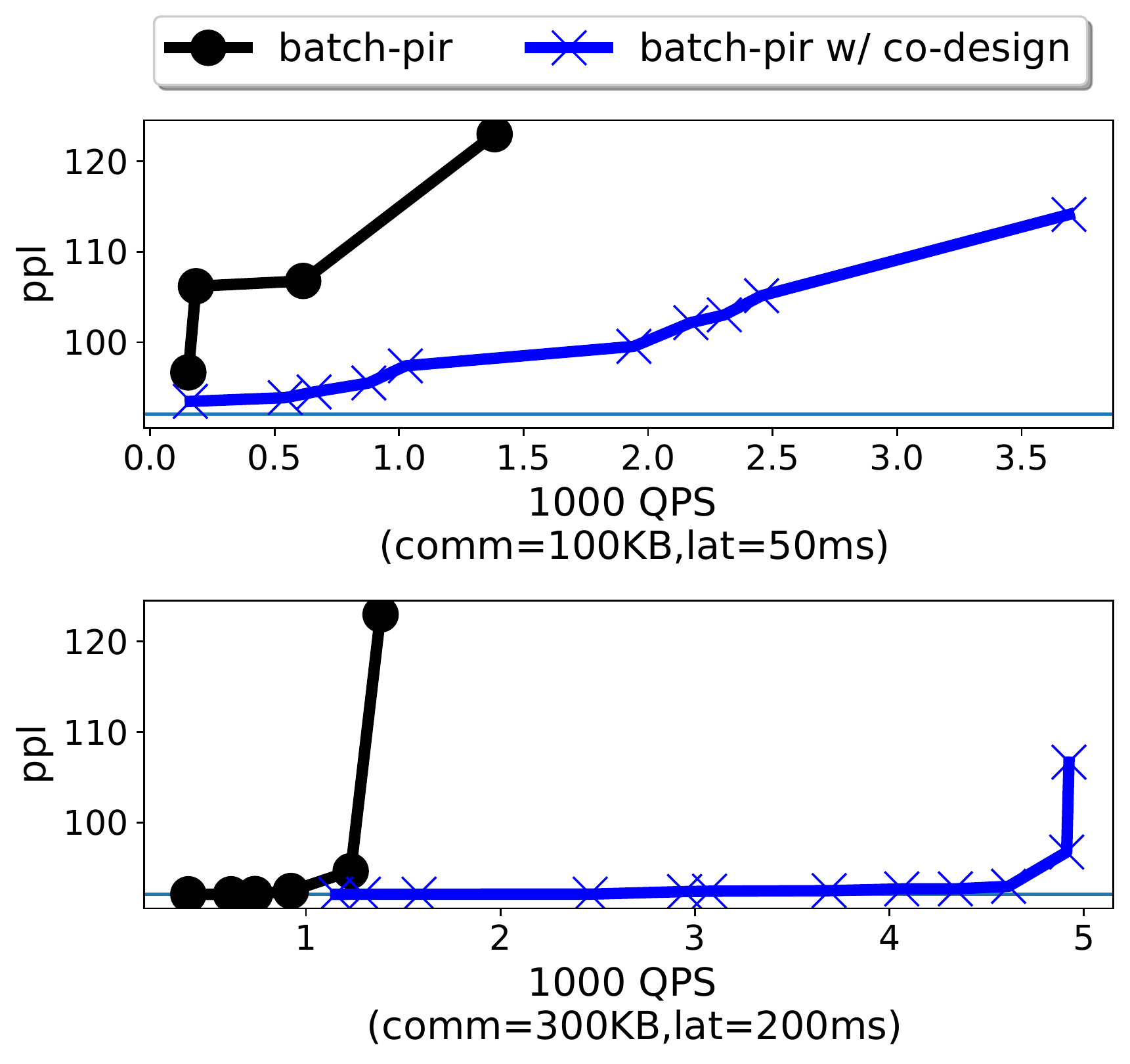} 
\vspace{-10pt}
\caption{System throughput vs model quality with and without co-design for language model across different budgets.}
\label{fig:tput_vs_acc_lm}
%\vspace{-5pt}
\end{figure}

\begin{figure}%[!h]
\centering
\includegraphics[width=.8\linewidth]{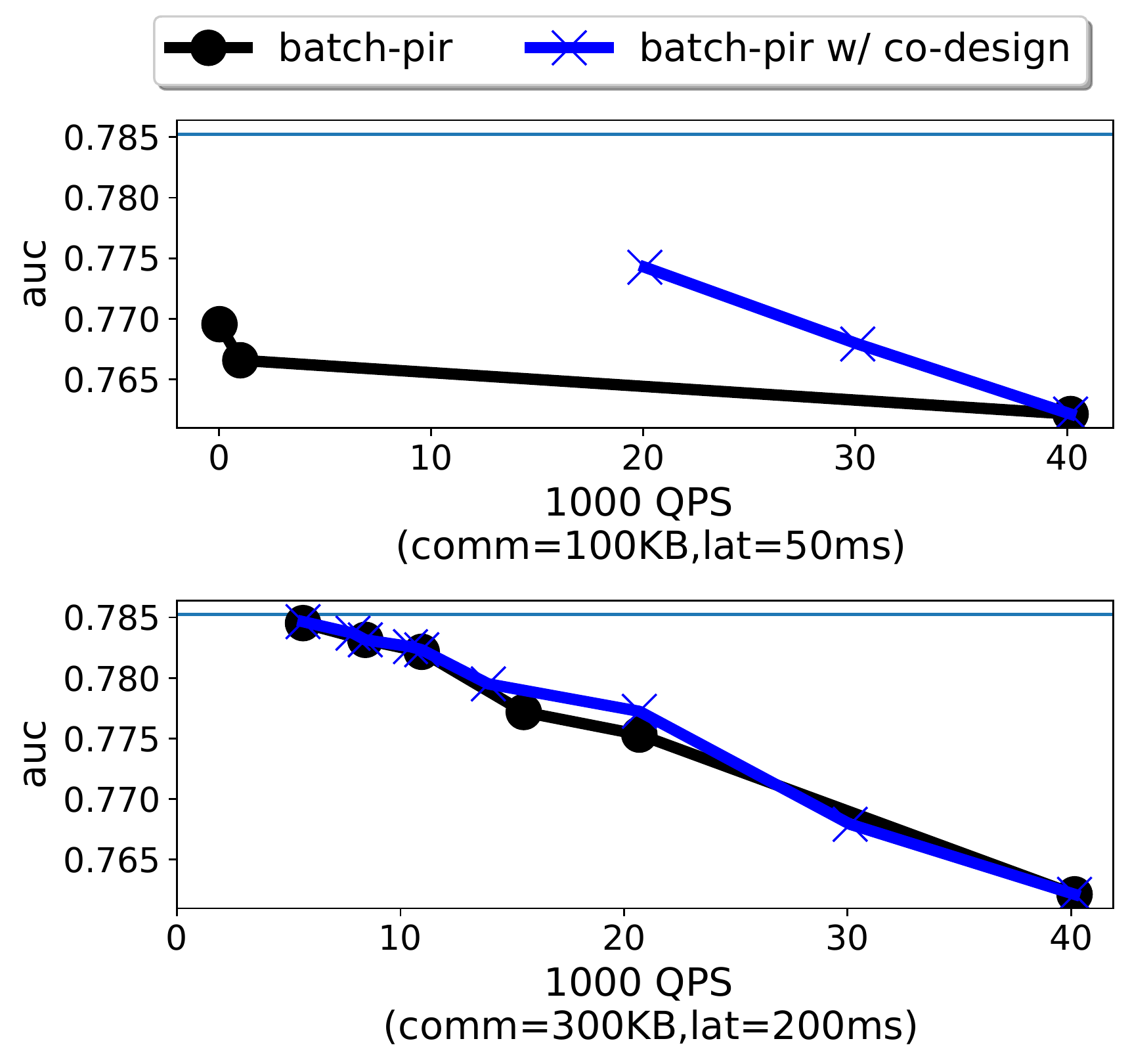} 
\vspace{-10pt}
\caption{System throughput vs model quality with and without co-design for MovieLens rec across different budgets.}
\label{fig:tput_vs_acc_movielens}
%\vspace{-5pt}
\end{figure}

\begin{figure}%[!h]
\centering
\includegraphics[width=.8\linewidth]{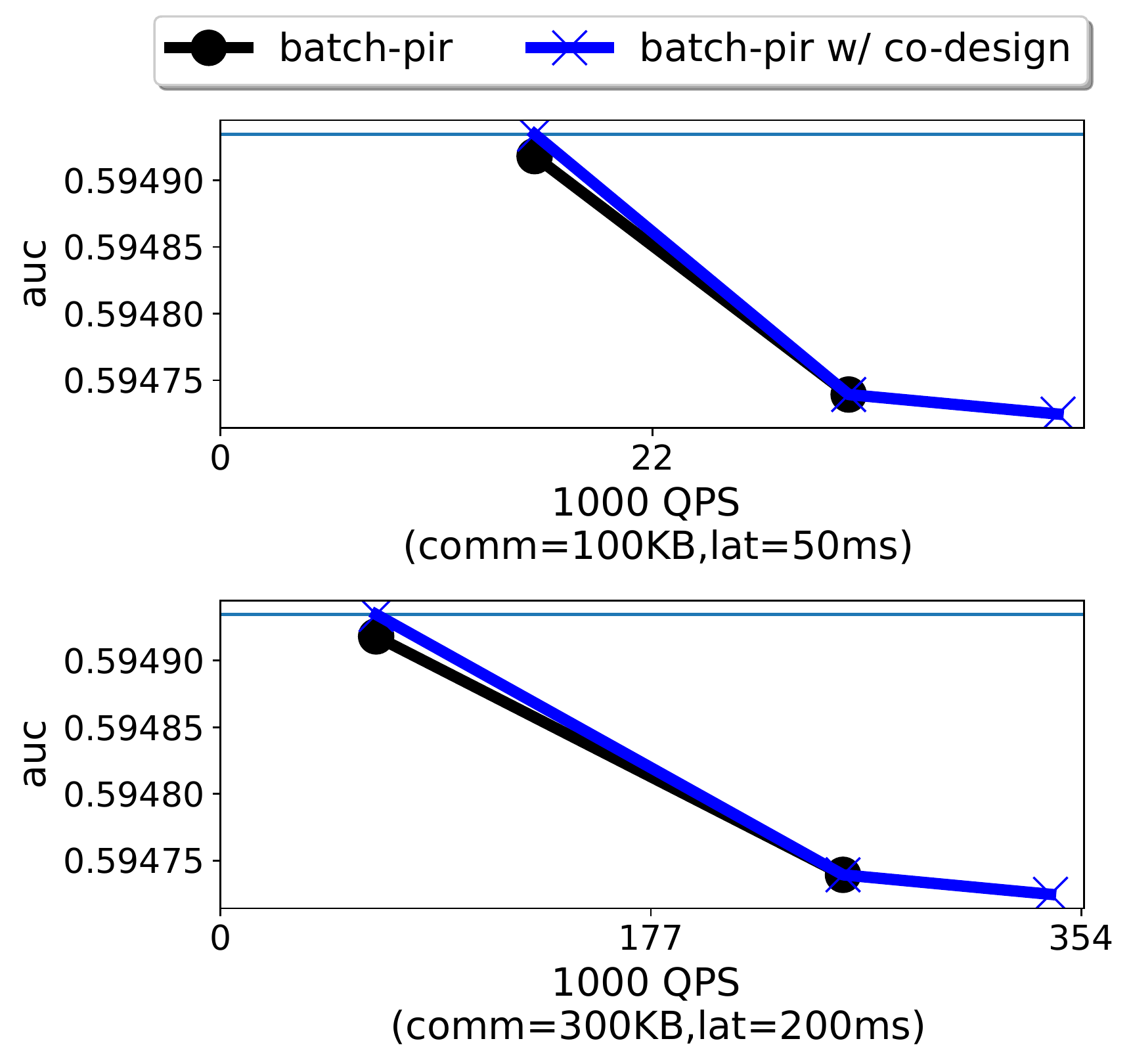} 
\vspace{-10pt}
\caption{System throughput vs model quality with and without co-design for Taobao rec across different budgets.}
\label{fig:tput_vs_acc_taobao}
%\vspace{-5pt}
\end{figure}

\noindent \textbf{Co-Design Throughput Improvement}
We show overall co-design throughput improvement over standard batch-PIR across all applications on select budgets in Figures \ref{fig:tput_vs_acc_lm}, \ref{fig:tput_vs_acc_movielens}, and \ref{fig:tput_vs_acc_taobao}. As shown, the PIR-ML co-design can result in significant improvements to the tradeoffs between model-quality and system throughput. 
Co-design is most effective when a) the budget is small enough to be sufficiently restrictive, and b) the impact of dropping queries has a significant impact on model quality. To expand on a), the budget plays a major role in the relative improvement that co-design sees as shown in Figures \ref{fig:tput_vs_acc_lm} and \ref{fig:tput_vs_acc_movielens}; there is increasingly smaller difference between batch-PIR and batch-PIR with co-design when the budgets are large enough. 
This makes intuitive sense as with a larger budget both batch-PIR schemes with and without co-design converge on the optimal pareto curve. 
Expanding on b), co-design is less helpful for applications where dropping the sparse features does not impact model quality -- this is natural since co-design optimizes for model quality and if the sparse features has less impact, the relative gains of co-design would also be less.
This phenomenon is best demonstrated by the observation that language model (Figure \ref{fig:tput_vs_acc_lm}) and MovieLens (Figure \ref{fig:tput_vs_acc_movielens}), whose model inputs are entirely sparse features that require embedding table lookups, see much greater improvement with co-design compared to Taobao (Figure \ref{fig:tput_vs_acc_taobao}), whose sparse categorical features are only a fraction of model inputs. 
Overall, the results show that PIR-ML co-design can significantly improve the system throughput beyond what just batch-PIR can support, especially under tight computation and/or communication budgets.

\section{Related Work}
\label{sec:related}
%\vspace{-0.05in}

\noindent \textbf{Privacy-preserving Computation Techniques} 
Prior work on privacy-preserving ML investigated techniques such as \rev{fully-homomorphic encryption} (FHE) \cite{gazelle, delphi}, \rev{secure multi-party computation} (MPC) \cite{crypten,cryptflow,falcon, ariann}, and trusted execution environments (TEEs) \cite{guardnn,mgx,secndp}. 
Unlike these prior studies, which primarily focus on protecting dense computation in neural networks, we investigate how to privately access large embedding tables in recommendation and language models. 

\rev{Recent work on FHE acceleration \cite{fab, heax, f1_he, bts, fx_henn, tensorfhe, poseidon, medha, sharp, cheetah, craterlake} suggests that FHE-based CNN models can run with low latency. Yet, they still suffer from low throughput. Due to the high computation demand of FHE, FHE accelerators typically use the entire chip (ASIC/FPGA/GPU) to run one inference at a time. 
%% ED: I'm commenting out the following as it feels too specific to call out one particular previous work. Originally, I added the number in the rebuttal to be more specific, but it seems less appropriate for the camera-ready.
%For example, ARK \cite{ark} is reported to be able to run 1 ResNet-20 inference in 125ms. This suggests that an ASIC can run 8 inferences per second. On the other hand, a NVIDIA H100 GPU is reported to be able to run 56,635 samples/second for a larger ResNet. 
While FHE has the potential to enable private inference for any model in the cloud, it is not yet efficient enough for high-throughput use~cases.} 

%ORAM~\cite{oram_orig,path-oram,fletcher2015freecursive,ren2013design} with a TEE or a trusted third-party will be another way to protect embedding table accesses on an untrusted cloud. Here, we use PIR in order to allow accesses from many client devices without any secure hardware or trusted third-party.

\noindent \textbf{Private Information Retrieval} 
PIR can be categorized into single-server protocols based on homomorphic encryption (HE) \cite{spiral, pir_he, pir_preprocess, pir_in_storage} and n-server (n $\ge$ 2) protocols based on DPFs \cite{blinder, riposte, floram}.
\rev{We focus on two-server DPF-based PIR protocols, as they are significantly more computation- and communication-efficient than single-server schemes~\cite{pir_in_storage, cheetah, bts, blinder, riposte, floram}.
For example, querying a 1B entry table with a two-server protocol is over 1000$\times$ more communication-efficient (2KB vs 3.6MB)~\cite{pir_in_storage} and multiple orders of magnitude more computationally-efficient than single-server protocols \cite{oneserverfortwo, cuckoo, mulpir, addra, onionpir, spiral}.}
\rev{
For a 1M-entry table, state-of-the-art HE PIR \cite{spiral} requires 14KB-60MB communication whereas our DPF-based system requires only 1.25 KB. 
HE PIR’s advantage over a DPF-based PIR system is that it only requires one server, rather than two non-colluding servers, enabling PIR under a stronger threat model. 
}
%compared to single-server HE based PIR \cite{pir_in_storage}, querying a 1B entry table costs only $2$KB of communication using DPFs, whereas using HE a query costs $2.6$MB, a $>$1000x overhead increase. Computation, similarly, is orders of magnitude heavier for HE based PIR, and requires HE operations versus PRF operations for DPFs; hence these approaches require custom hardware to be practical \cite{pir_in_storage}}. 
 Compared to n$>$2 DPF approaches, two-server DPF-based PIR protocols are more communication-efficient: 2-server DPF exhibits $O(log(n))$ communication ~\cite{dpf_1, dpf_2} while n${>}$2-server DPF exhibits $O(\sqrt{n})$ communication ~\cite{dpf_3}.

%\rev{Compared to single-server PIR schemes based on homomorphic encryption (HE) \cite{spiral}, DPF-based PIR schemes as employed in this paper require considerably less communication and computation. Concretely, on a 1M-entry table, state-of-the-art HE PIR requires 14KB-60MB communication \cite{spiral} whereas our DPF-based system requires only 1.25 KB. HE PIR’s advantage over a DPF-based PIR system is that it only requires one server, rather than two non-colluding servers, enabling PIR under a stronger threat model. 
%}

%\noindent \textbf{Feasibility of the Non-colluding Assumption} 
The two-server PIR protocols require the two participating servers hosting the (embedding) tables to be non-colluding.
This threat model with two (or more) non-colluding servers is commonly used in a large body of work on secure multi-party computation (MPC)~\cite{crypten,cryptflow,falcon,ariann,blinder,riposte,floram}.
Different from other computation with MPC, in DPF, no communication is required between the servers, and thus, the two servers can be hosted by different cloud providers with minimal performance overhead.  
Further, recent advances in MPC platforms make such a system increasingly realistic~\cite{mpc_alliance, fb_mpc, ACC, amazon_tee, forbes_mpc, google_mpc}. One realistic scenario is \rev{for the companies that want to provide strong privacy standards to} form a consortium to act as each others' non-colluding second party; these efforts \cite{mpc_alliance, meta_privacy_enhance} are seeing increasing adoption.
Remote attestation capabilities in public cloud TEEs \cite{amazon_tee,ACC,gcp_tee,secndp} can also be used to further ensure the integrity of two parties. 

\noindent \textbf{Batch Private Information Retrieval}
Various approaches for batch PIR \cite{multihot, cuckoo, batch_codes, cuckoo, poly_codes} have been proposed. We show that noise tolerance of ML allows the use probabilistic PIR protocols like \cite{multihot} with minimal accuracy loss.

\noindent \textbf{On-device ML}
On-device ML has been studied for recommendation \cite{on_device_rec, edgerec}, speech recognition \cite{amazon_on_device_speech}, translation \cite{on_device_translation}, etc. 
Our work uses on-device ML for privacy, and enables the private use of large server-side embedding tables.

%applications that access an embedding table that may be too large to store practically on device. on-device recommendation models have begun to become popular due to recent privacy regulations, and embedding tables remain a system challenge despite attempts to compress them. Our work proposes to use PIR to overcome these privacy and system challenges.

\section{Conclusion}
%\vspace{-0.05in}
We present a system for efficiently and privately serving embeddings for on-device ML application.  Our system on a single V100 GPU can serve up to $100,000$ queries per second---a ${>}100\times$ speedup over naive system, enabling practical deployment for privacy-sensitive applications.

%%%%%%%%%%%%%%%

%\bibliographystyle{ACM-Reference-Format}
\bibliographystyle{acm}
\bibliography{references}

\appendix

\end{document}